\documentclass[reprint,float fix, superscriptaddress, preprint numbers, nofootinbib]{revtex4-2}
\usepackage[]{graphicx}
\usepackage{epstopdf}
\usepackage{array}
\usepackage{amsmath}
\usepackage{subfigure}
\usepackage{subcaption}
\usepackage{bm}
\usepackage{hyperref}
\usepackage{framed}
\usepackage{MnSymbol}
\usepackage[dvipsnames]{xcolor}
\usepackage{ulem}
\usepackage{float}
\usepackage{dcolumn}
\usepackage{bm}
\date{\today}

\begin{document}
\title{Evidence of Athermal Metastable Phase in a Halide Perovskite: Optically Tracked Thermal-Breach Memory}
\author{Kingshuk Mukhuti}\email{Joint first-authors with equal contribution}
\affiliation{Indian Institute of Science Education and Research Kolkata, Mohanpur, Nadia 741246, West Bengal, India}
\author{Satyaki Kundu}\email{Joint first-authors with equal contribution}
\affiliation{Indian Institute of Science Education and Research Kolkata, Mohanpur, Nadia 741246, West Bengal, India}
\author{Debasmita Pariari}\email{Joint first-authors with equal contribution}
\affiliation{Solid State and Structural Chemistry Unit, Indian Institute of Science, Bengaluru 560012, India}
\author{Deepesh Kalauni}
\affiliation{Indian Institute of Science Education and Research Kolkata, Mohanpur, Nadia 741246, West Bengal, India}
\author{Ashutosh Mohanty}
\affiliation{Solid State and Structural Chemistry Unit, Indian Institute of Science, Bengaluru 560012, India}
\author{Aniket Bajaj}
\affiliation{Indian Institute of Science Education and Research Kolkata, Mohanpur, Nadia 741246, West Bengal, India}
\author{D. D. Sarma}\email{sarma@iisc.ac.in}
\affiliation{Solid State and Structural Chemistry Unit, Indian Institute of Science, Bengaluru 560012, India}
\author{Bhavtosh Bansal}\email{bhavtosh@iiserkol.ac.in}
\affiliation{Indian Institute of Science Education and Research Kolkata, Mohanpur, Nadia 741246, West Bengal, India}
\date{\today}
\begin{abstract}
Halide perovskite materials have been extensively studied in the last decade because of their impressive optoelectronic properties. However, their one characteristic that is uncommon for semiconductors is that many undergo thermally induced structural phase transitions. The transition is hysteretic, with the hysteresis window marking the boundary of the metastable phase. We have discovered that in methylammonium lead iodide, this hysteretic metastable phase is athermal, meaning it shows almost no temporal phase evolution under isothermal conditions.  We also show that a large number of distinguishable metastable states can be prepared following different thermal pathways. Furthermore, under a reversible thermal perturbation, the states in the metastable phase either show return-point memory or undergo a systematic nonrecoverable phase evolution, depending on the thermal history and the sign of the temperature perturbation. Since the phase fraction can be probed with extreme sensitivity via luminescence, we have an optically retrievable memory that reliably records any breach in temperature stability. Such thermal-breach memory in athermal martensites, of which there are numerous examples, may be useful for tagging packages requiring strict temperature control during transportation or preservation.
\end{abstract}
\preprint{Phys. Rev. Lett. (2025)}
\maketitle
The physics of condensed matter has traditionally been concerned with matter close to thermal equilibrium where the observables are history-independent single-valued functions of the state variables \cite{Chaikin}. Nevertheless, there has been continued interest in transcending these equilibrium constraints and exploring complex matter with static memory \cite{Stern-Murugan, Keim_RMP, Nova_SrTio3, Bensea_PNAS, Ordonez-Miranda, Mungan_PNASReview, Pershin-DiVentra, Jiang_FerroelectricPRL2023}.  Ferromagnetic hysteresis is, of course, a historically familiar example  \cite{Kronmuller,   Mayergoyz_2003Book, Bertotti, Barker_ProcRoySoc1983} but more recently memory in a variety of athermal mechanical systems---from corrugated sheets \cite{Bensea_PNAS} to sheared granular matter \cite{Candela2023_GranularMemory, Reichhardt_2023}---is being extensively studied within the thermodynamics framework.

A necessary condition for matter to have memory is that the equilibrium characteristics of history independence and single valuedness be relaxed. For such memory to be useful, one would further require the material to possess systematic and reproducible behavior.
\begin{figure}[!b]
	\includegraphics[scale=0.195]{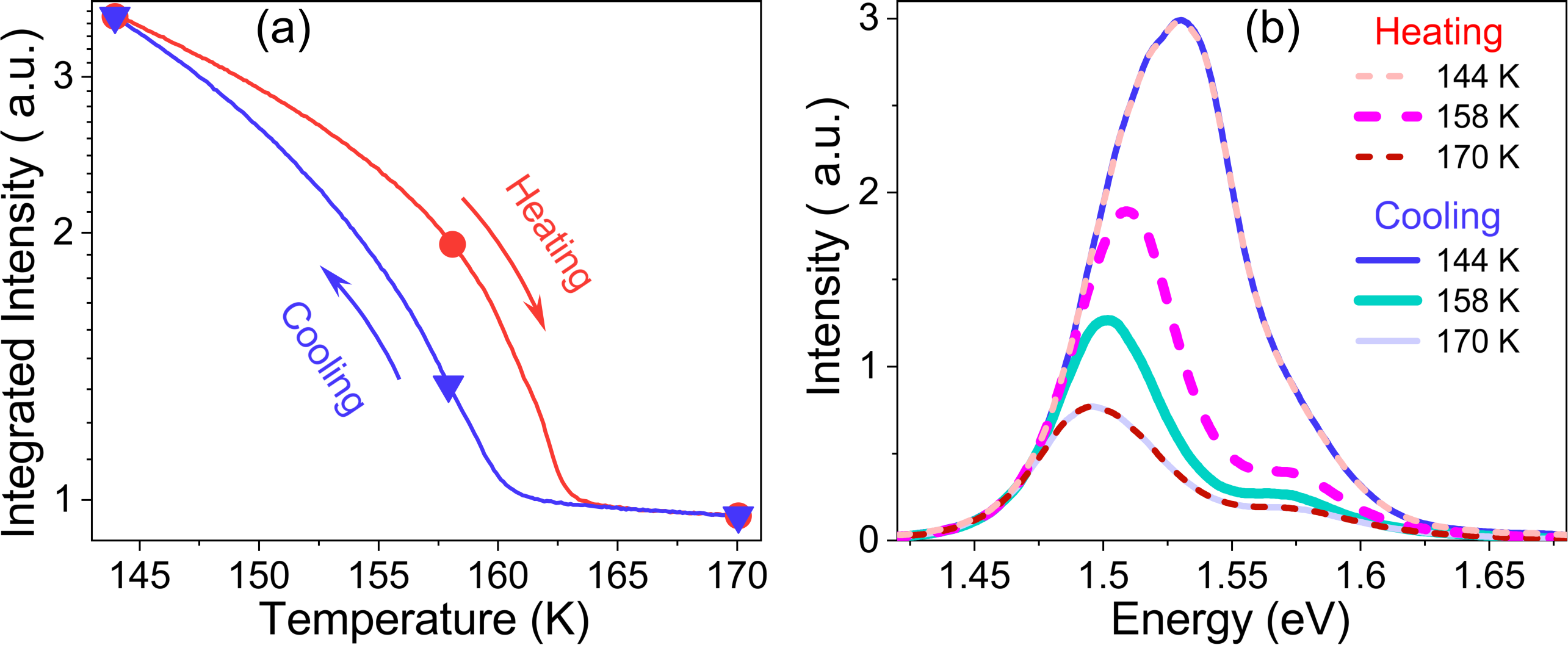}
	\centering
	\caption{Abrupt phase transition and metastability in MAPbI$_3$. (a) Temperature dependence of the integrated photoluminescence (PL) intensity tracks the strongly hysteretic phase transition. The red curve corresponds to heating and the blue to cooling. (b) A few representative PL spectra between $\sim$ 144-165 K. Note that the spectra taken in the metastable phase at 158 K are strongly history dependent.}
	\label{Fig1}
\end{figure}

In this Letter, we report the discovery of unusual metastable states that are spontaneously created around the thermally induced phase transition in methylammonium lead iodide (MAPbI$_3$). Such halide perovskite semiconductors have garnered significant attention from the optoelectronics community on the account of ease of growth, tolerance to defects, high luminescence yield \cite{Akkerman, Hutter_MaPbI3, MAPBI3_Structural, Abdi-Jalebi_MAPbI3, Sharada,Brenner_MAPbI3, Ashutosh_2019}, and other unusual characteristics \cite{Fabini-emphanisis, Kang_JPCL_2021, Debasmita_JACS2023, Debasmita_ACSEL2024}.  As is typical of halide perovskites \cite{Kang_JPCL_2021, Lin-Lai_Dou}, MAPbI$_3$ undergoes a structural phase transition with prominent thermal hysteresis. We have discovered that the states in the hysteretic region show deterministic features. This is quite at variance with what is expected for a first-order transition undergoing classical nucleation \cite{Lin_VO2_PRL2022, Chaikin} and instead similar to the phenomenology of martensites \cite{Ortin_Planes_Delaey, Bhadeshia, Nishiyama}.

The phase transition, despite being thermally induced is {\it athermal}. By athermal, we mean that temperature has little thermodynamic significance and rather acts like a dynamical parameter, at par with strain, pressure, or magnetic field \cite{Perez-Reche_PRL2001, Chandni, Nandi}. The metastable states show a near absence of evolution if the temperature is held fixed, suggesting a negligible role of diffusion in the phase transformation \cite{Bhadeshia, Nishiyama}. One can create (and destroy) a hierarchy of states in a systematic and apparently deterministic manner. This athermal character implies that the materials exhibit a {\it thermal-breach memory}.

The evolution of the photoluminescence (PL) spectra of MAPbI$_3$  \cite{Suppl} across its first-order orthorhombic-to-tetragonal structural phase transitions around 160 K \cite{MAPBI3_Structural, Lin-Lai_Dou, Ashutosh_2019} are summarized in Fig. 1. The transition is captured by the sharp drop in the PL intensity and a thermal hysteresis loop enclosing the region of metastability where the PL intensity is multivalued in temperature.

\begin{figure}[!h]
	\includegraphics[scale=0.18]{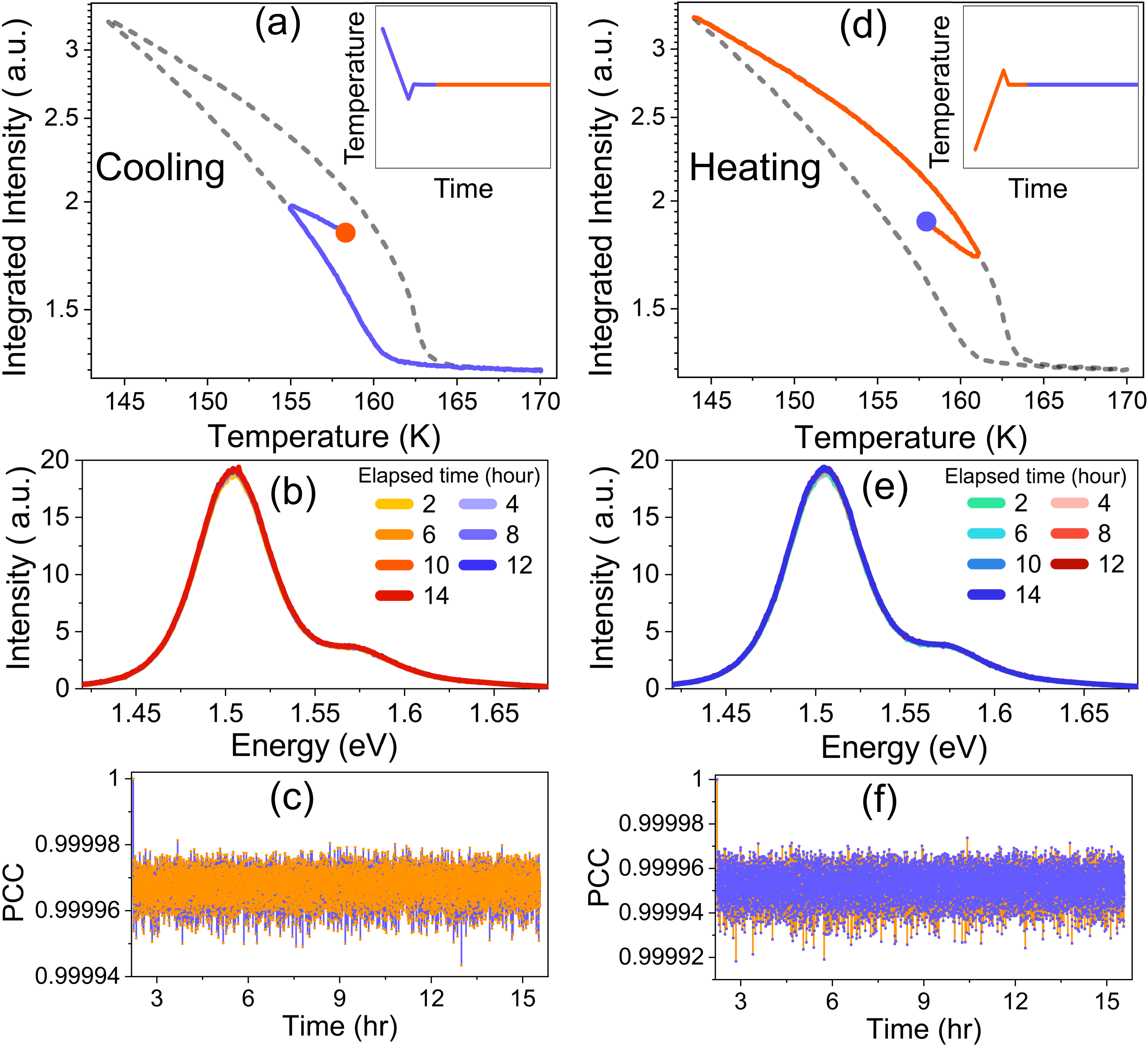}
	\centering
	\caption{Arrested metastability in MAPbI$_3$. Two metastable states with similar integrated PL intensity (phase fraction) are prepared at 158 K following complementary thermal protocols, as shown in panels (a) and (b), discussed in detail in the text. The part of the major hysteresis loop that is not traversed by the sample is shown by dotted lines. (c),(d) show the PL spectra measured over a period of 12 h at a fixed reservoir temperature of 158 K. (e),(f) The Pearson correlation coefficient (PCC) between the first and the subsequent spectra over this 12-h period. The origin of the time axis is taken to be the start of the experiment and the data are shown here from 2 h onward, the time for state preparation and temperature stabilization to $\pm 0.005$ K.}  
\label{Fig2}
\end{figure}

{\em Nonevolving athermal metastable phase}---Figure 2 illustrates the temporal stability of the coexisting phases. A metastable state is created by traversing a partial thermal loop [170 K $\rightarrow$ 155 K$\rightarrow$ 158 K], as shown in Fig. 2(a) (inset). The PL intensity through this process is traced (blue line) in Fig. 2(a). After reaching the target temperature of 158 K, the sample is held fixed at $158.000 \pm 0.005$ K for many hours and the PL spectra are monitored [Fig. 2(b)].

The changes between the spectra measured at the time $t_0$ and subsequent times $t_k$ may be quantified by the Pearson correlation coefficient (PCC) $C({t_0, t_k})$. 
\begin{equation}
	C({t_0, t_k})= {\int_{\varepsilon_i}^{\varepsilon_f} d\varepsilon\, \textrm{PL}(t_0, \varepsilon) \times \textrm{PL}(t_k, \varepsilon) \over \sqrt{\int_{\varepsilon_i}^{\varepsilon_f} d\varepsilon\, [\textrm{PL}(t_0, \varepsilon)]^2}\sqrt{\int_{\varepsilon_i}^{\varepsilon_f} d\varepsilon\, [\textrm{PL}(t_k, \varepsilon)]^2}}.
\end{equation}

Here, $\textrm{PL}(t, \varepsilon)$ is the PL intensity at time $t$ and energy  $\varepsilon$. The $\varepsilon_i$ and $\varepsilon_f$ are the lower and the upper limits of the energy where the PL emission is significant. Figure 2(c) shows that $C({t_0, t_k})$, with \relpenalty=10000 $t_k\in [t_0, 14 \,\textrm{h}]$ is constant within  $2\times 10^{-5}$. The degree of the arrest \cite{fn_defects} of the kinetics can be assessed by comparison with the correlation coefficient's value of $\sim 0.7$ between the two spectra at 144 and 170 K in Fig. 1(b) at the two ends of the metastable region.

In Figs. 2(d)-2(f), we repeat the above experiment by creating a different metastable state, also at 158 K, but now starting from the low-temperature side and by following a complementary thermal path [144 K $\rightarrow$ 161 K $\rightarrow$ 158 K]. Note the parallel between the two columns of Fig. 2. Remarkably, as we will see below, these two states respond very differently to thermal perturbation.
\begin{figure}[!h]
	\includegraphics[scale=0.265]{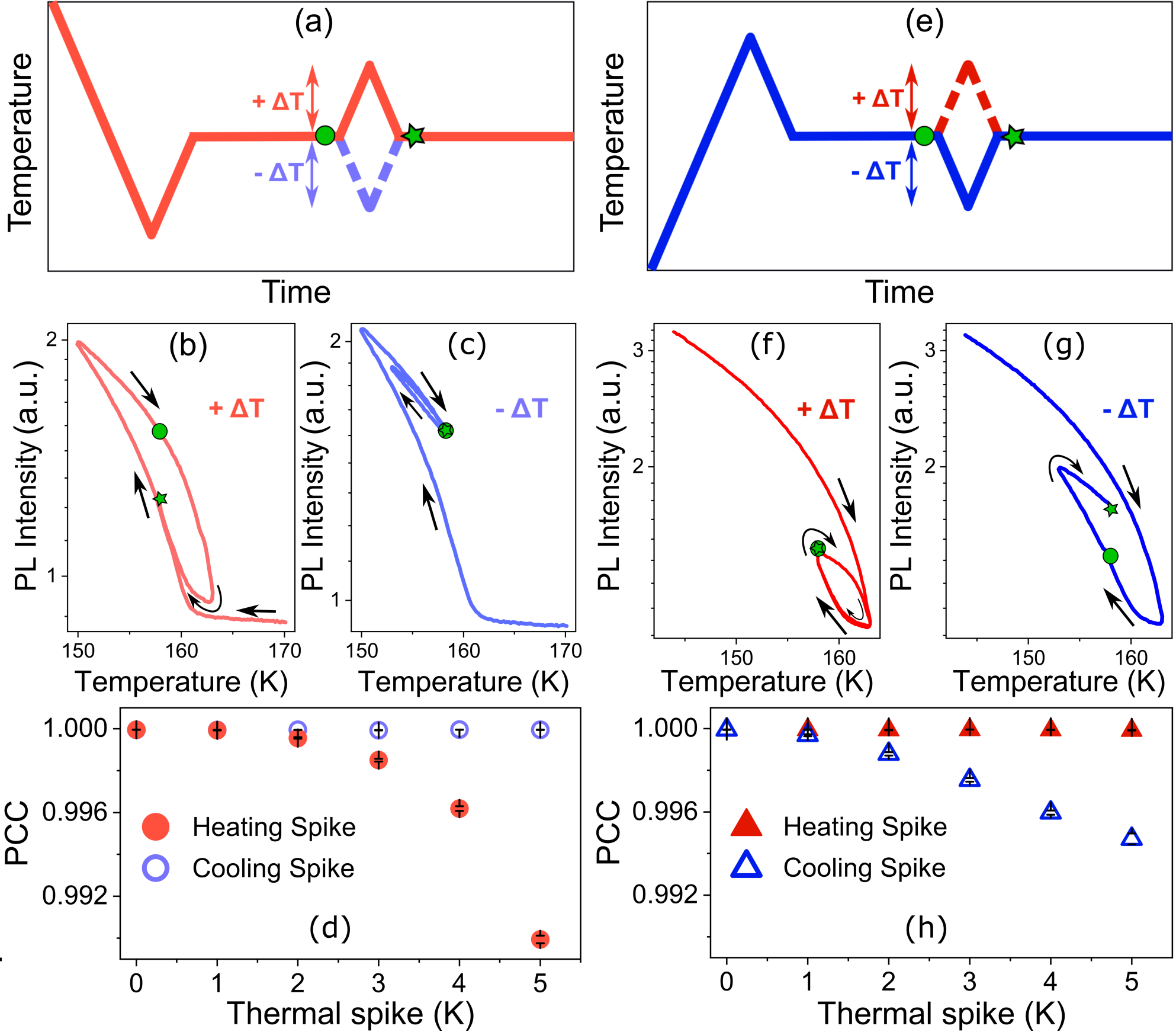}
	\centering
	\caption{Sensitivity of the metastable phase to thermal perturbation---thermal breach memory. Following Fig. 2, two different metastable states of nearly the same phase fraction are prepared at the same temperature $T_0=158$ K by minor hysteresis loops traversed in the opposite sense. In (a) the temperature $T_0$ is reached with the final step involving an increase in temperature, whereas in (e) the temperature was decreased to reach the target temperature $T_0$. After waiting for 20 min, the sample is made to undergo a small reversible thermal perturbation $\Delta T$ to reach the final state, state $\pentagram$. The beginning of the thermal perturbation is marked as state $\circ$. The changes in the integrated PL intensity (for $\Delta T=\pm 5$ K) are shown in (b) and (c) for the protocol in (a), (f), and (g) for the protocol in (e), respectively.   For the thermal history (a), only the heating thermal spike ($+\Delta T$) causes a nonrecoverable phase evolution [as seen in (b)], and for the thermal history (e), only the cooling thermal spike ($-\Delta T$) causes a non-recoverable phase evolution [as seen in (g)]. A recoverable evolution is observed in the other two cases shown in (c) and (f). Panels (d), (h): the experiment is now repeated with different $\Delta T$ values. PCC of the spectra between state $\circ$ and state $\pentagram$. The wiping out is understood in terms of the return-point memory property that is discussed in Fig. 5 below.} 
	\label{Fig3}
\end{figure}

{\em Thermal-breach memory}---Let a state be prepared following the path 170 K$\rightarrow$ 150 K$\rightarrow$ 158 K. Let us call this state $\circ$. State $\circ$ which is at 158 K is now subjected to a reversible thermal perturbation $\Delta T$ and returned back to the same temperature of 158 K. Let us call the system to be now in state $\pentagram$ [Fig. 3(a)]. The experiment is repeated ten times for $\Delta T=\pm 1, \pm 2, \pm 3, \pm 4, \pm 5$ K.
The effects of this brief thermal perturbation on the PL intensity are shown in Figs. 3 (b) and 3 (c) for the positive and negative $\Delta T= 5$ K, respectively. With $-\Delta T$, state $\circ$ and state $\pentagram$ are identical [Fig. 3(c)], while state $\pentagram$ appears distinctly different for a positive $\Delta T$ [Fig. 3(b)]. Thus, the PL intensity, via its dependence on the phase fraction, detects any positive temperature fluctuation even when the temperature returns to the original $T$. The PCC between the spectra from the state $\circ$ and state $\pentagram$ cooling and heating spikes is shown in Fig. 3(d).

Figures 3(e)-3(h) describe the experiment when the state at 158 K is prepared following the complementary thermal protocol (144 K$\rightarrow$ 163 K $\rightarrow$ 158 K). The phase evolution is now nonrecoverable for $-\Delta T$ and recoverable for $+\Delta T$. The extent of recovery can also be quantified by the PCC. For a thermal spike of $-5$ K the PCC between the state $\pentagram$ and the state $\circ$ is $0.99992 \pm 1.44 \times 10^{-5}$ [Fig. 3(d)] and for $+5$ K it is  $0.99994 \pm 9.2 \times 10^{-6}$  [Fig. 3(h)]. These values are of comparable magnitude to the values in Figs. 2 (c) and 2 (f); the sample while still in the metastable phase behaves as if it never experienced the 5 K perturbation. This spectacular recovery is the manifestation of the more general return-point memory phenomenon  \cite{Keim_RMP, Bertotti, Sethna_RPP, Bensea_PNAS, Ortin_Planes_Delaey,    Mayergoyz_2003Book}. For cases where a nonrecoverable evolution is observed, a temperature spike of $\Delta T=1$ K has a change of about $10^{-3}$ in the PCC. This is easily discernible.

\begin{figure}[h!]
	\begin{center}
		\includegraphics[scale=0.46]{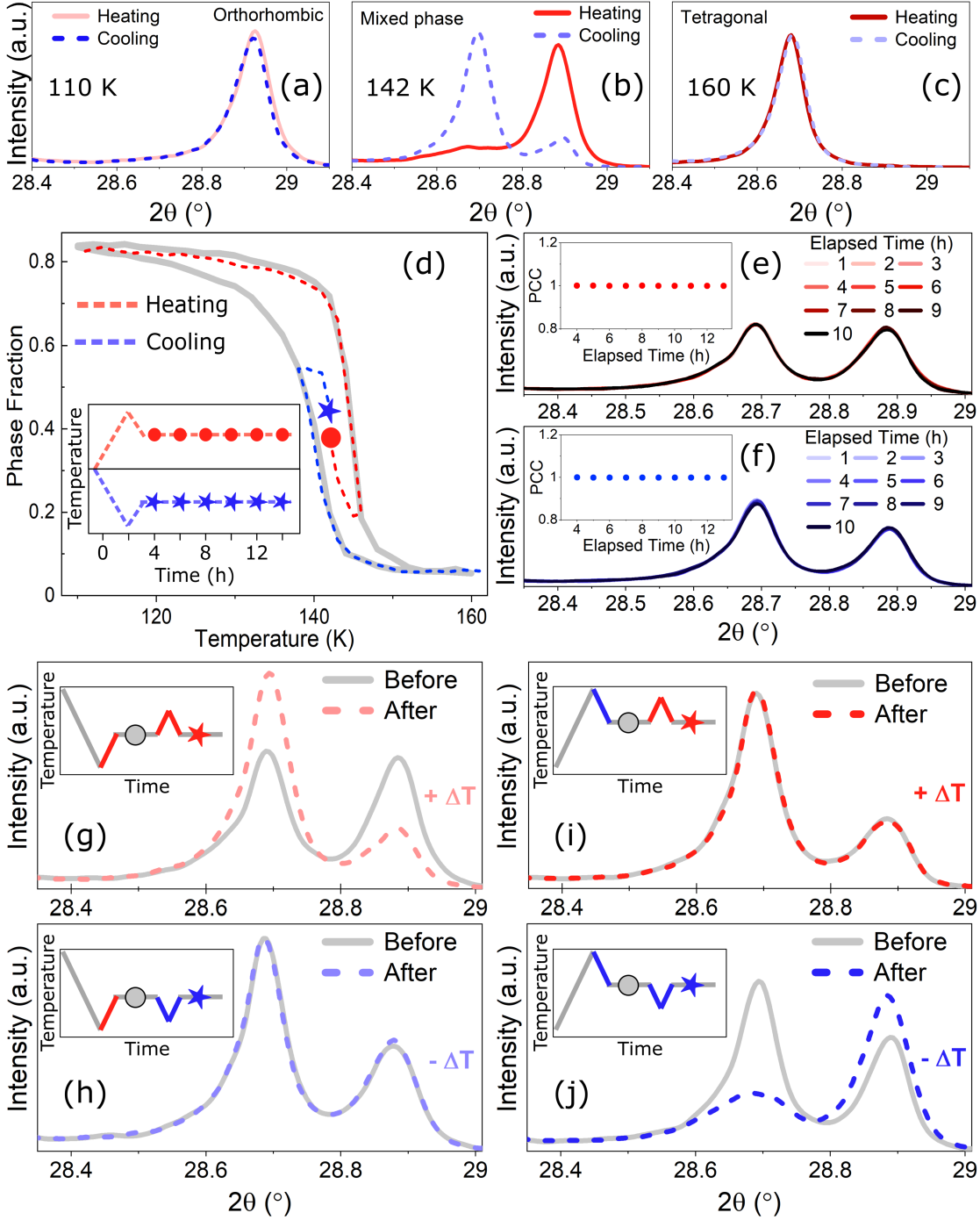}
		\caption{{Powder XRD measurements to independently establish the athermal phase transition and thermal breach memory \cite{Suppl}}. The $2\theta$ peaks (a) at 28.9$^\circ$ and (c) at 28.7$^\circ$ are from the low-temperature orthorhombic and the high-temperature tetragonal phases, respectively; with the metastable mixed phase showing both peaks in panel (b). (d) The gray solid line traces the hysteresis revealed by the phase fraction determined from the XRD results. Following the two temperature protocols shown in the inset, the system evolves to the metastable states $\pentagram$ and $\circ$, respectively. (e),(f) Absence of phase evolution, captured through both nonevolving $2\theta$ peaks as well as the near-unity PCC, establishes the athermal character, consistent with Fig. 2. The manifestation of the thermal breach memory [in panels (g) and (j)], and the recoverability [in panels (h) and (i)] is directly demonstrated by the phase fractions, providing a microscopic understanding of the corresponding results obtained from PL experiments (compare Fig. 3).  Note the slight difference in the transition temperatures on account of the difference between growth runs.}
	\end{center}
	\label{Figure:rpm_main}
\end{figure}

{\em X-ray diffraction measurements}---While the PL spectra capture the system’s macroscopic evolution well, XRD can directly measure the orthorhombic-to-tetragonal phase fraction, which may be considered the order parameter of this phase transition. We find that the previous assertions based on the PL measurements are perfectly corroborated by powder x-ray diffraction (XRD) measurements on MAPbI$_3$. The reflections associated with (202) planes for the low-temperature orthorhombic phase and (220) planes for the higher-temperature tetragonal phases exhibit distinct $2\theta$ peaks [Figs. 4(a)-4(c)], allowing for direct estimation of the phase fraction [Fig. 4(d)]. Figures 4(d)-4(f) closely parallel Fig. 2 and establish the arrested phase evolution in the metastable phase under isothermal conditions and thus the athermal character of the transition. Figures 4(g)-4(j) show the thermal breach memory effect, paralleling Fig. 3. 

{\em Return-point memory}---Return-point memory indicates the existence of a hierarchy of states in the metastable region. Let us create such a hierarchy by making temperature reversal cycles described in Fig. 5(a). The resulting integrated PL intensity shows a sequence of nested loops [Fig. 5(b)]. At the first temperature reversal at 155 K, marked state 1, the system exits the major hysteresis loop and starts on a path inside the metastable region. Minor loops formed by thermal reversals at 156 K [state 2] and 157 K [state 3] are found to be contained within the bigger loops. Finally, starting at state 3, we make one last minor loop (see figure caption) where the temperature is first raised to 160 K, and from this inflection point, lowered all the way down to 100 K. In the journey from the extremum of the innermost of the nested loops at 160 K, we find that the material retains the memory of the past thermal history, and state 3, state 2, and state 1 are sequentially crossed.  Furthermore, on reaching back to state 2 at 157 K, the memory of the smallest excursion (157 K $\rightarrow$ 160 K$\rightarrow$ 157 K) is lost and memory of the second minor loop  (156 K $\rightarrow$ 161 K$\rightarrow$ 156 K) is wiped out below 156 K. Finally, on crossing 155 K, the system joins the major hysteresis loop and memory of the first minor loop (155 K $\rightarrow$ 162 K$\rightarrow$ 152 K) is also wiped out. A completely symmetric behavior is observed under reverse thermal cycling \cite{Suppl}. This behavior is also found in the phase fractions determined from equivalent x-ray diffraction experiments \cite{Suppl}.

\begin{figure}[h!]
	\begin{center}
		\includegraphics[scale=0.2]{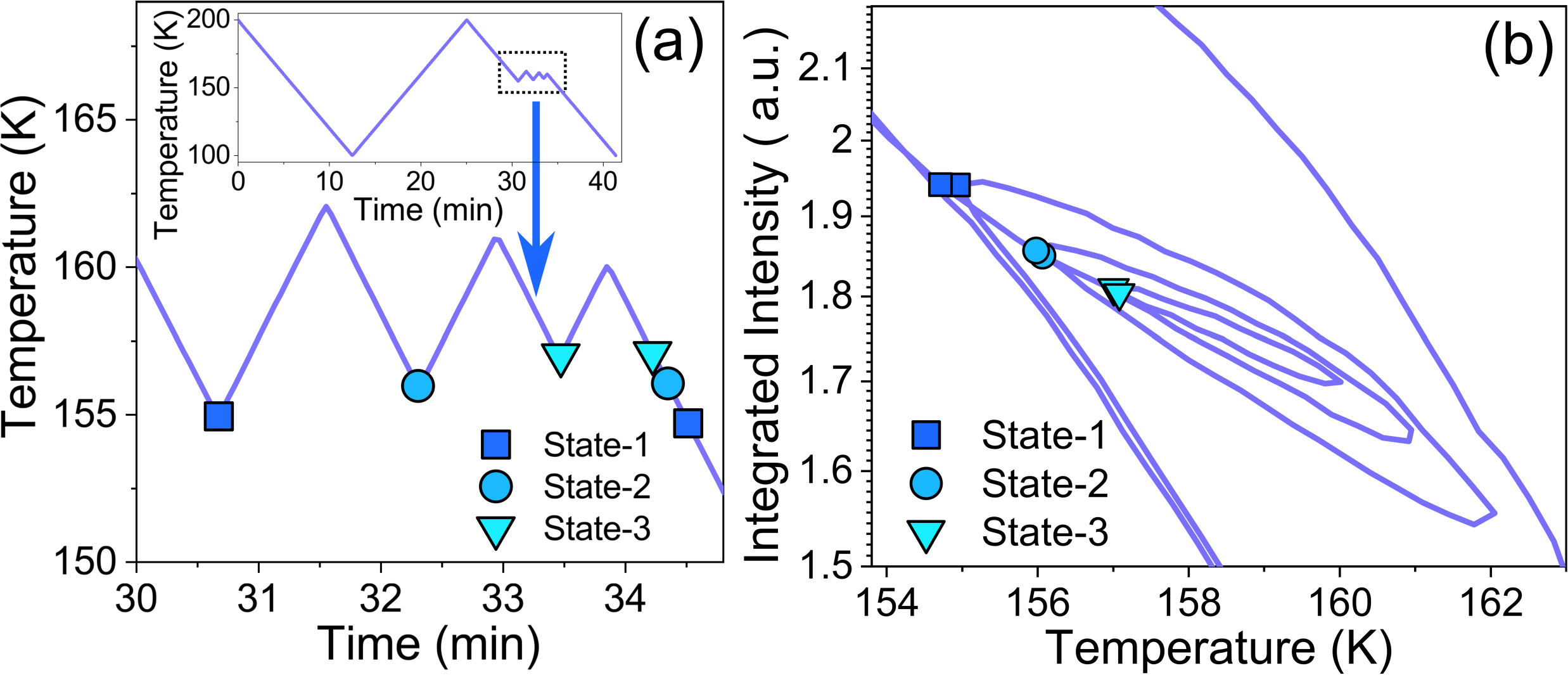}
		\caption{{\it Return-point memory in the metastable phase.} (a) Starting from the high-temperature phase at 200 K, the temperature is cycled through 200 K $\rightarrow$ 155 K $\rightarrow$ 162 K $\rightarrow$ 156 K $\rightarrow$ 161 K $\rightarrow$ 157 K $\rightarrow$ 160 K $\rightarrow$ 100 K.  (b) This results in a sequence of nested hysteresis loops of the PL intensity. (a),(b) At the first thermal inflection point at 155 K, marked as state 1, the system exits the major hysteresis loop and starts a minor loop (155 K $\rightarrow$ 162 K $\rightarrow $ 156 K). Before this loop can be completed, there the temperature is reversed at 156 K. This point is marked as state 2. Similarly, a second incomplete minor loop of smaller amplitude (156 K $\rightarrow$ 161 K$\rightarrow$ 157 K) is now made. The point at 157 K is marked state 3. A still smaller loop is now started at 157 K, with the inflection temperature at 160 K.  Finally, sitting within this smallest hysteresis loop at 160 K,  we lower the temperature to 100 K. As the temperature is lowered the system systematically passes through state 3, state 2, and finally state 1 to join the major hysteresis loop.}
	\end{center}
	\label{Figure:rpm_main}
\end{figure}

{\em Discussion}---The phase transition in MAPbI$_3$ has many curious features. These are not reconciled within the usual understanding of the first-order phase transition, where hysteresis and metastability are necessarily kinetic phenomena \cite{Debenedetti}, related to a delay on account of the nucleation barriers. For the first-order transitions, nucleation is a diffusion-governed stochastic phenomenon that is unlikely to show the strict reversibility manifest in the return-point memory.

On the other hand, the athermal phenomenology is reminiscent of the pinning-induced effects in the charge-density-wave transitions \cite{Wang_Ong, Sethna_RPP, Keim_RMP} or the behavior seen in martensites and shape-memory alloys \cite{Ortin_Planes_Delaey, Perez-Reche_PRL2001, Gottschall, Olemskoi-Katsnelson}, and complex correlated oxides \cite{Satyaki_PRL, Ordonez-Miranda}. We also note the remarkable similarity with hysteretic ferromagnetism \cite{Suppl}. 

Return-point memory \cite{  Mayergoyz_2003Book, Barker_ProcRoySoc1983}, in particular, suggests that the system is amenable to being modeled as a collection of Preisach-type hysterons \cite{Mayergoyz_2003Book, Bertotti,  Mungan_PNASReview,Keim_RMP}. This is demonstrated in the Supplemental Material where we interpret the return-point memory experiment of Fig. 5 in terms of systematic switching of hysterons. The Preisach measure is determined in a separate experiment by the first-order-reversal curves measurement \cite{  Mayergoyz_2003Book}.

Despite this useful representation of the metastable phase via the Preisach model, we emphasize that the statistical mechanics of athermal behavior at finite temperature is not quite understood. While the local mechanical equilibria between the incompatible low and high-temperature phases can be well explained within the nonlinear elasticity theory \cite{Zhang_James_Muller, Bhattacharya}, the associated thermodynamic theory of the transition itself belongs to the relatively nascent field of statistical mechanics of nonadditive long-range interactions \cite{Latella_Madrid_Ruffo}, with the strain being that long-range interaction controlling the transition \cite{Gagne_Gould_Klein}. Although spin models with disorder---the random-field Ising model \cite{Sethna_RPP} and the Sherrington-Kirkpatrick spin-glass model \cite{Pazmandi_PRL1999}---can reproduce these features of hierarchical organization and return-point memory, the treatments have been limited to zero temperature. 

Unlike the magnetic systems where the long-range dipolar fields of the ferromagnetic domains can perhaps suppress the thermal fluctuations, the control variable for us is not the magnetic field, but the temperature itself. Thus, paradoxically, MAPbI$_3$ has a thermally induced (entropy-driven)  phase transition that also behaves ``athermally" (Fig. 2), viz., the temperature acts merely as a parameter without causing diffusion. 

Temperature cycling can also deterministically prepare the system to encode a hierarchy of states. These states can, in principle, be used for information storage \cite{Perkovic_Sethna}. How much information? Roughly, assuming that the hysteresis window is $\approx 20$ K [Fig. 1] and that stable athermal states can be defined to a resolution of $1$ K, we have $N=20$ unique temperature set points in our hysteresis window. It is then not hard to show that a small macroscopic volume of the sample can then have $[2^{20}-1] \approx 10^6$ configurable states that are accessible via different sequences of temperature reversals \cite{Suppl}. These configurations are not independently addressable bits of conventional memory but irreducibly combine into hierarchical structures. Such an ability to encode hierarchy is one hallmark of complexity \cite{ladyman}. 

Furthermore, the arrested phase evolution at a fixed temperature (Fig. 2), combined with that of a phase evolution accompanying a small temperature change (Fig. 3), also suggests another potentially important memory application, quite different from the much-discussed resistive and neuromorphic memories \cite{Kang_JPCL_2021}. Under many situations, including those involving biological samples, materials must be transported or stored under conditions of stringent thermal control. A mixed metastable state of an athermal luminescent material (prepared as a powder or paint) can serve as an optically accessible memory; any breach in temperature control (e.g., thaw and recooling, or its inverse thermal spike, that are simulated in Fig. 3) is recorded via a nonrecoverable phase evolution and a corresponding change in the emission spectrum. We have further shown that PCC [Figs. 3(d) and 3(h)] with respect to the original spectrum makes a sensitive and reliable measure of the breach. Moreover, PCC does not require the knowledge of the absolute PL intensity and thus, can easily be computed from spectra taken at different locations by different spectrometers.

While the work is not on single crystals and does not elucidate the actual mechanism of the athermal transition with regards the microscopic long-range force balance that suppresses diffusion,  these results do highlight the rich and systematic character of metastability in the hysteretic phase change region of MAPbI$_3$. The fact that the material also happens to be highly luminescent in both phases is a happy coincidence that additionally allows for the phase fraction to be probed by noninvasive and easy-to-operate optical means.

{\em Acknowledgments}---S. K. and D. P. thank the Council of Scientific and Industrial Research (CSIR), India for financial support. B. B. thanks the Science and Engineering Research Board (SERB), Department of Science and Technology, Government of India, for the Core Research Grants (No. CRG/2018/003282 and No. CRG/2022/008662). D. P., A. M., and D. D. S. thank the Science and Engineering Research Board (SERB), and the Department of Science and Technology, Government of India for support of this research. D. D. S. thanks Jamsetji Tata Trust, J. C. Bose Fellowship of SERB, and CSIR Bhatnagar Fellowship for their support.

\newpage
\newpage
\onecolumngrid


    \setcounter{figure}{0}
    \setcounter{subfigure}{0}
	     \setcounter{section}{0}
		\begin{center}
        $********$
        \vspace{1cm}
        
			{\Large {\bf Supplemental Material}}
		\end{center}
		\noindent

		In this Supplemental Material, we discuss some background details regarding the preparation and characterization of the sample, and elaborate on some other measurements that present additional supporting observations; especially about the slow relaxation and return point memory phenomenon. The hysteretic metastable phase is also discussed using the Preisach model.  		
\tableofcontents
	\section{Sample synthesis and characterization}
We synthesized polycrystalline methylammonium lead iodide (MAPbI$_3$) through a two-step solution route. Initially, we prepared methyl ammonium iodide (MAI) by mixing 15.5 mL (180 mmol) of methylamine and HI precursors and stirring the mixture at 0$^\circ$C for 2 h. Subsequently, we performed a distillation at 60$^\circ$C to remove the excess HI. we then repeatedly washed the resulting crude product with diethyl ether until it turned completely white. Afterward, we recrystallized it in ethanol. After repeated washing of the recrystallised product with diethyl ether we dried the white solid MAI under vacuum and stored it in an inert atmosphere.
	
We synthesized MAPbI$_3$ by dissolving 0.1589 g (1 mmol) of MAI and 0.461 g (1 mmol) of PbI$_2$ in 1 mL of $\gamma$-butyrolactone to create a solution with a 1 M concentration with respect to the lead content. We stirred the mixture at 55$^\circ$C for 30 minutes, resulting in a bright yellow solution. They subsequently drop-cast the solution onto a clean glass slide preheated to 125$^\circ$C and annealed it at the same temperature for 20 min to evaporate the solvent. Finally, we scraped off and collected the resulting gray-black shiny MAPbI$_3$ crystalline powder for further measurements, including PXRD, which revealed no impurity peaks. We characterize them with the PXRD measurements carried out on a {\it PANalytical Empyrean} diffractometer at 45 kV, 30 mA, under Cu-K$\alpha$ radiation ($\lambda$ = 1.54059 \AA). The acquired XRD data is further analyzed using {\it PANalytical X'Pert HighScore Plus} software and we depict the diffraction pattern in Fig. 1[supplement]. The profile fitting is performed using {\it FullProf\_Suite} software. The profile-fitted red line, which is in good agreement with the experimental pattern in Fig. 1[supplement], ensures that the sample is devoid of any impurities.
\begin{figure}[h]
\centering
\includegraphics[scale=0.5]{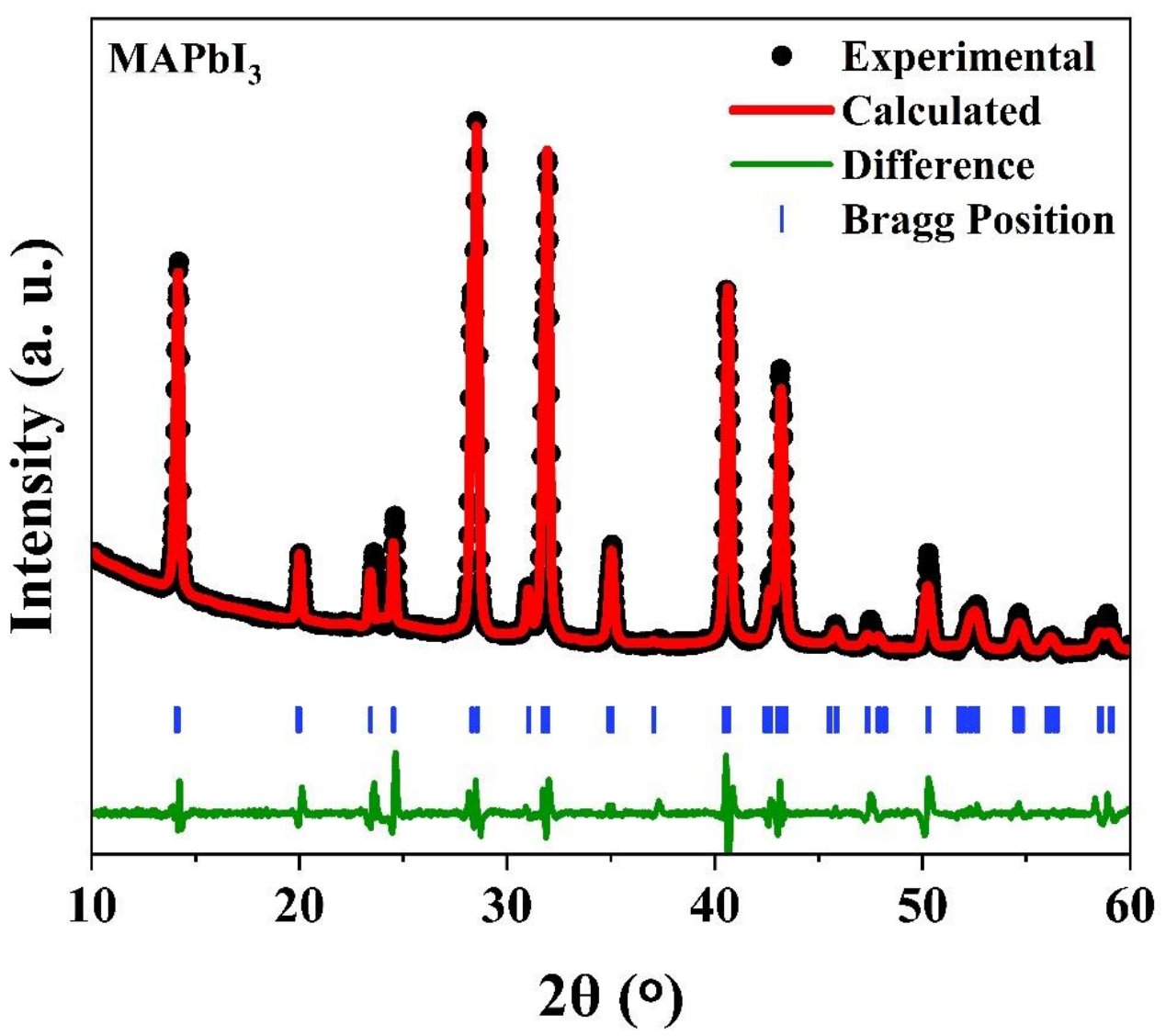}\hspace{0em}%
\caption{[supplement] The room temperature PXRD of a typical MAPbI$_3$ sample along with the profile fitting. The residual of the fit is also shown.}
		\label{fig:T_2}
	\end{figure}
\section{Temperature-dependent XRD measurements}
The PXRD measurements were carried out on a {\it Bruker D8 Discover} diffractometer at 40 kV and 40 mA, under Cu-K$\alpha$ radiation ($\lambda$ = 1.54059 \AA). Low-temperature PXRD measurement was performed using the same diffractometer, coupling it with {\it Oxford Cryosystem} closed-cycle helium cryostat. PXRD data analyses were done using {\it PANalytical X'Pert HighScore Plus} software.  We have used standard XRD data from the Inorganic Crystal Structure Database (ICSD), Crystallography Open Database (COD), and Cambridge Structural Database (CSD) for comparison.
\begin{figure}[H]
\begin{center}
\includegraphics[scale=0.6]{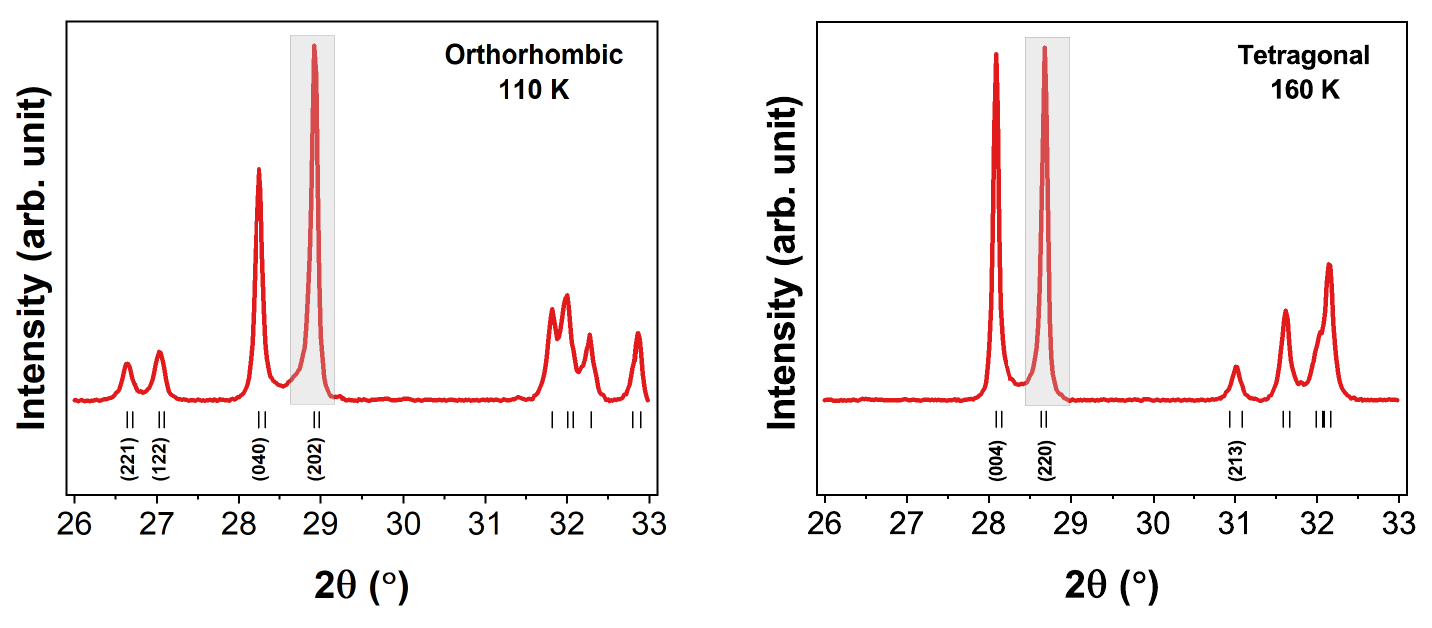}
\caption{(a) [supplement] Temperature evolution of the $2\theta$ peaks of the PXRD spectrum used to track and characterize the metastable phase in Fig. 4[main text]. The PXRD spectrum in the range $2\theta\in [26^\circ,33^\circ]$ in the low-temperature orthorhombic phase and the high-temperature tetragonal phase has distinct differences which can be used to track the phase evolution. For this purpose, we have traced the evolution of the peaks associated with the (202) plane in the orthorhombic phase to the (220) plane in the tetragonal phase. This is highlighted in gray.}
\end{center}
\end{figure}

\setcounter{figure}{1}
\begin{figure}[H]
\begin{center}
\includegraphics[scale=0.45]{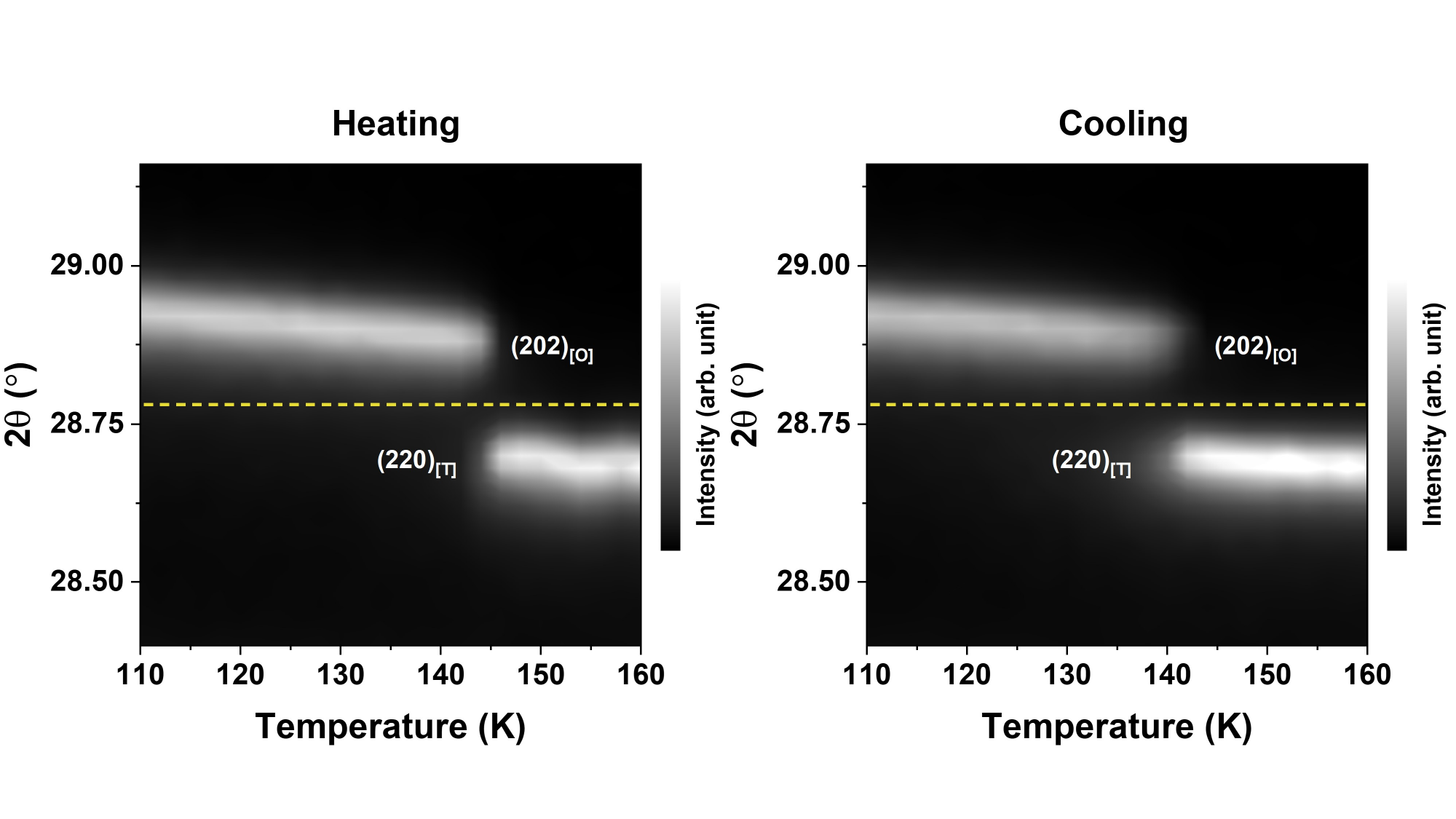}
\caption{(b) [supplement] Temperature evolution of the $2\theta$ peaks of the PXRD spectrum used to track and characterize the metastable phase in Fig. 4[main text]. Pseudo-color intensity maps show the evolution of the peaks associated with (220) plane in the tetragonal phase and (202) plane in the orthorhombic phase around $2\theta = 28.75^\circ$. This evolution was used to track the phase transition and characterize the mixed phase in the metastable phase in Fig. 4[main text].  Note the slight variation in the phase transition temperature for the samples grown in different batches.}
\end{center}
\end{figure}

Figure 2(b)[supplement] presents two graphs, illustrating the diffraction patterns during heating and cooling, respectively. The graphs depict the changes in diffraction intensity as a function of temperature, ranging from 110 K to 160 K. A dashed yellow line at 28.75$^\circ$ is used as a reference to separate the diffraction peaks, acknowledging a slight overlap between them. The intensity of the peaks is measured in arbitrary units. 

The integrated intensities of the diffraction peaks are used to determine the phase fraction. The orthorhombic-to-tetragonal phase fraction [Fig. 4(main text)] is calculated as 
\[
\text{Phase fraction} = \frac{I_{[O]}}{I_{[O]}+I_{[T]}}
\]
where $I_{[O]}$ is the integrated intensity of the (202)$_{[O]}$ peak and $I_{[T]}$ is the integrated intensity of the (220)$_{[T]}$ peak.
\section{Photoluminescence measurements}
	The sample was excited by a 489 nm continuous wave diode laser coupled to a fiber optic cable in a home-built liquid nitrogen cryostat. The cryostat was optimized for flexibility regarding temperature cycling and provided excellent temperature control (Fig. 3[supplement]) between 77 K to 350 K.  The temperature of the sample could be varied with a linear ramp rate from 0.1 K/min to 30 K/min [e.g., Figs. 2(a), 2(d)[main text]], and in this range of values, the transition temperature and the PL spectra were independent of the temperature ramp rate.
 
	\begin{figure}[h]
		\centering
		\includegraphics[clip, scale=0.3]{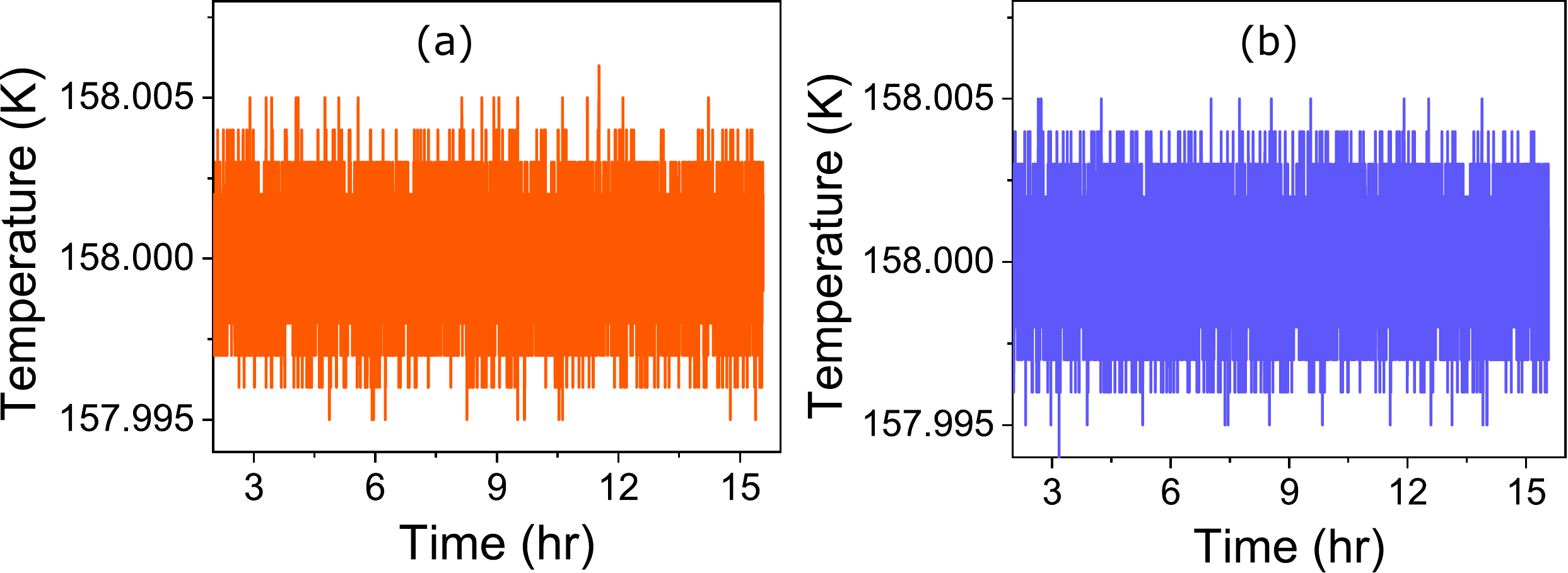}
		\caption{[supplement] The temperature stability of the cryostat while recording the PL spectra for the metastable states at 158 K. The final state is prepared through two similar but opposite thermal protocols (shown in Figs. 2(a), 2(d)[main text]) before we stay there for more than 12 h, and the destination point is reached by (a) heating and (b) cooling. The $\pm 5$ mK fluctuation in both cases is essentially limited by the digitization of the temperature controller.}
		\label{fig:T_2}
	\end{figure}
	
	In order to probe the optical response of MAPbI$_3$ in and out of the metastable region of the orthorhombic$\leftrightarrow$tetragonal phase transition we excite the sample with a 489 nm continuous wave diode laser inside a fiber-coupled liquid nitrogen cryostat and record PL spectra at every 3 s. Such a protocol is implemented without heating the sample by modulating the laser, synchronous to an EMCCD with an `on' time of 4 ms. 
	
	\section{Estimation of spectral dissimilarity} In this work we use the Pearson correlation coefficient (PCC), defined in Eq. 1(main text) to quantify the dissimilarity between any two PL or XRD spectra. Since the value of this correlation coefficient has no absolute meaning, we highlight the magnitude of the variation of the PCC across the hysteretic region. These numbers provide a useful comparison with the values of the PCC reported in Fig. 2[main text] and Fig. 3[main text].
 \begin{figure}[h]
	  \begin{center}
			\includegraphics[scale=0.3]{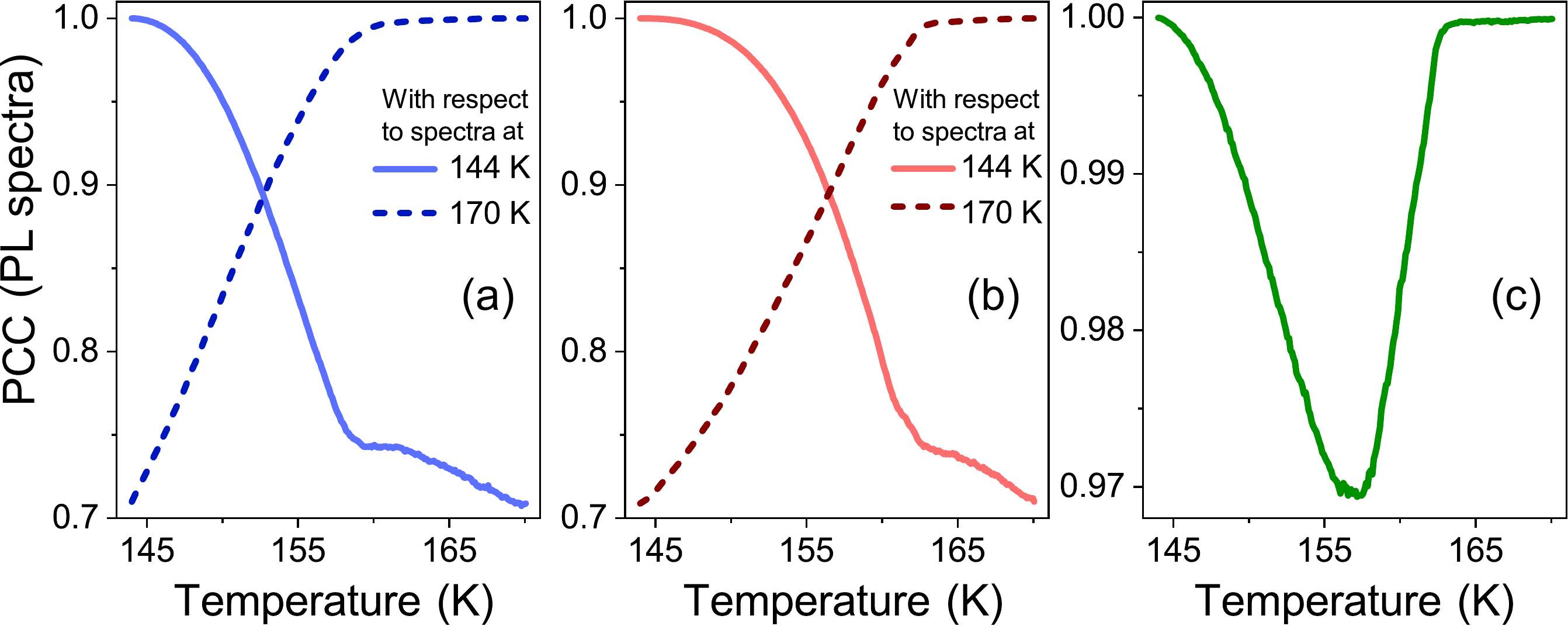}
			\caption{[supplement] The dissimilarity between the PL spectra in the low-temperature orthorhombic and the high-temperature tetragonal phase quantified by the PCC. (a),(b) Changes in the PL spectra, and thus the extent of phase evolution, are quantified by the correlation (PCC) between the PL spectra in one of the pure phases (at 144 K or 170 K) and at an intermediate temperature. (a) PCC of all spectra during a cooling cycle with respect to the spectra at 144 K (solid) and 170 K (dashed). (b) PCC of all spectra during a heating cycle with respect to spectra at 144 K (solid) and 170 K (dashed). (c) PCC between the spectra at the same temperature during heating and cooling cycles to highlight the hysteretic metastable phase. PCC value is nearly 1 after 165 K and before 144 K, implying that heating and cooling spectra in these regions are very similar. The deviation from 1 in the range 144 K-165 K is because of the dissimilarity of the spectra at the same temperature measured during heating and cooling, which again highlights metastability and hysteresis in this region.}
		\end{center}
  \vspace{0.5 cm}
 \end{figure}
 \begin{figure}
		\begin{center}
			\includegraphics[scale=0.3]{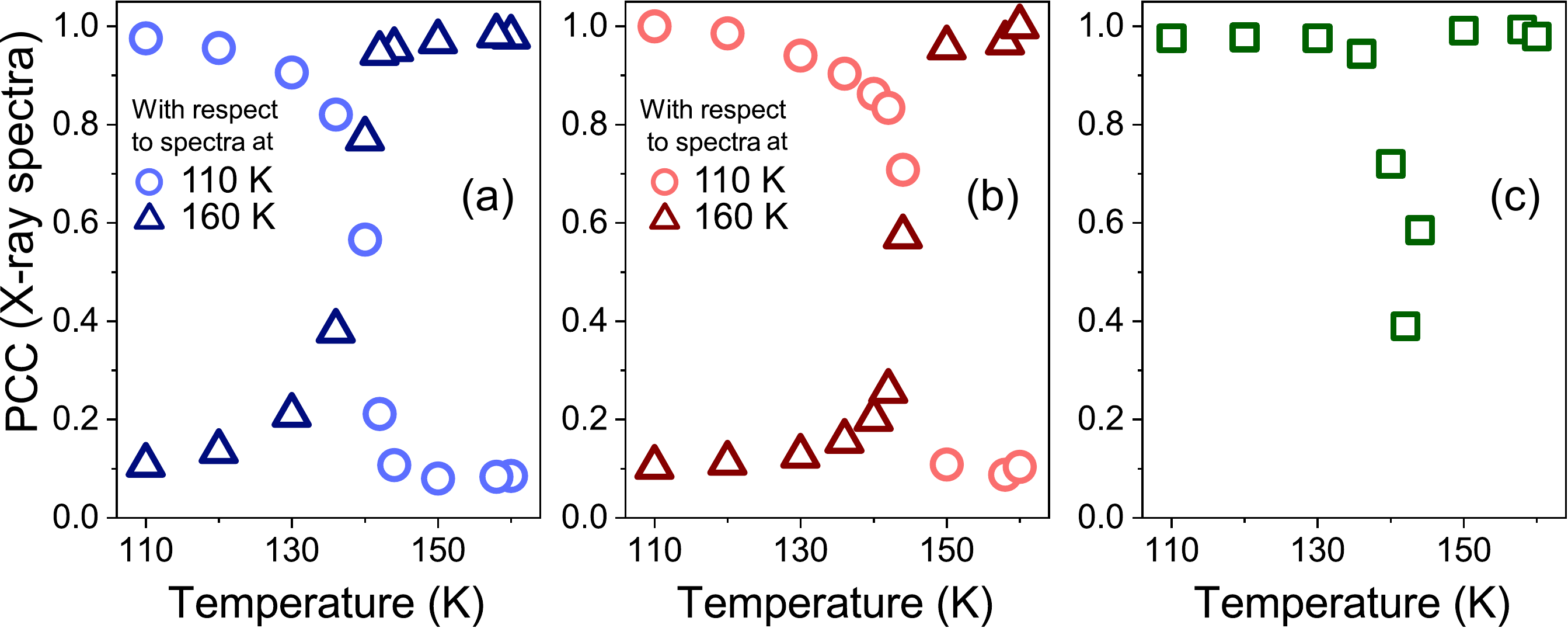}
			\caption{[supplement] The dissimilarity between the powder XRD spectra from the (202) planes in the low-temperature orthorhombic and the high-temperature tetragonal phase quantified by the PCC. (a)-(c) closely follow the three panels above but note that the transition temperature for this sample is slightly different from the transition temperature for the sample where the PL is measured.}
		\end{center}
 \end{figure}
Figures 4(a)-4(c)[supplement] quantify the evolution of the PL spectrum through the phase transition. In Fig. 4(a)[supplement], the two curves depict the PCC values for the spectra, measured between 144 K-170 K, during the cooling run correlated with the spectra measured at the two extreme ends of the heating cycle, viz, at 144 K and 170 K respectively. Figure 4(b)[supplement] is just the complement of Fig. 4(a)[supplement] and we now show the PCC calculation where the spectra measured during the heating run are correlated with the spectra at 144 K and 170 K, respectively. Therefore, Figs. 4(a), 4(b)[supplement] show that the PCC drops down to a minimum value of about 0.7, depicting the maximum dissimilarity between the spectra at two extremes of this hysteretic transition, irrespective of whether the heating or cooling protocol is chosen. On the other hand, Fig. 4(c)[supplement] portrays the dissimilarity between the spectra acquired at the same temperature during heating and cooling cycles, showing that the width of the hysteresis is maximum at around 158 K, having minimum spectral PCC of 0.97.

Figure 5[supplement] closely parallels Fig. 4[supplement]. Here we plot the PCC between the XRD spectra from the (202) planes.
\clearpage

\section{Return point memory--Supplementary data}
In this section, we will present further data and supporting information on the return-point memory experiment of the photoluminescence intensity described in Fig. 5[main text]. 

Figure 6[supplement] is the companion of Fig. 5[main text]. Here the return-point memory phenomena is studied for the complementary thermal protocol, with the minor loop excursions made on the heating part of the major hysteresis loop.  
\begin{figure}[h] \label{figsupp:RPP_Heating}
		\begin{center}
			\includegraphics[scale=0.4]{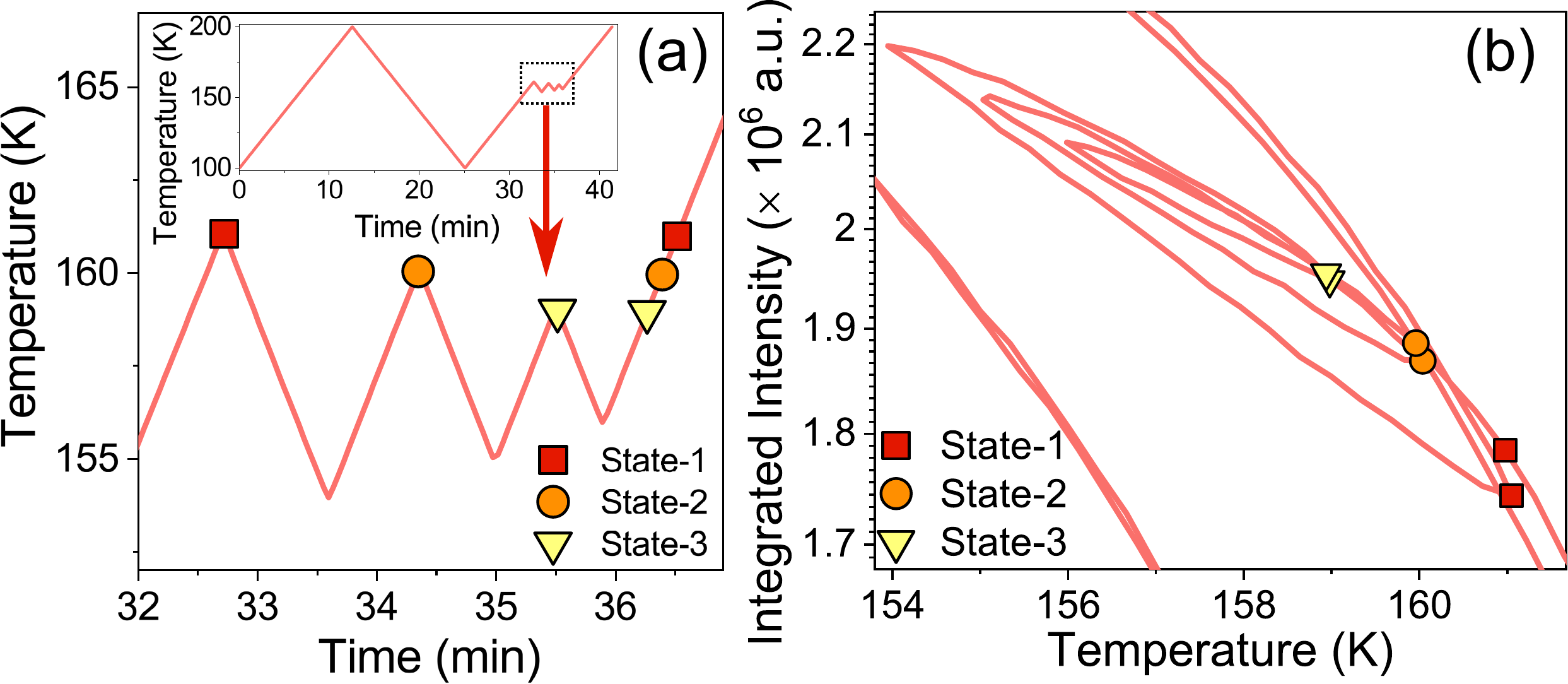}
			\caption{[supplement] Return-point memory in PL (heating loop). This figure closely follows Fig. 5[main text] but is done under a complementary thermal protocol. (a) Temperature cycling to form minor hysteresis loops with the three inflection points at 161 K, 160 K, and 159 K marked as state 1, state 2, and state 3, respectively. Inset shows the temperature variation through the entire experiment and shows the location of these minor loops on the major loops made between 100 K$\rightarrow$ 170 K$\rightarrow$100 K$\rightarrow$ 170 K. (b) The corresponding PL intensity.}
		\end{center}
	\end{figure}
	\begin{figure}[h!]\label{figsupp:RPP_spectra}
		\begin{center}
			\includegraphics[scale=0.3]{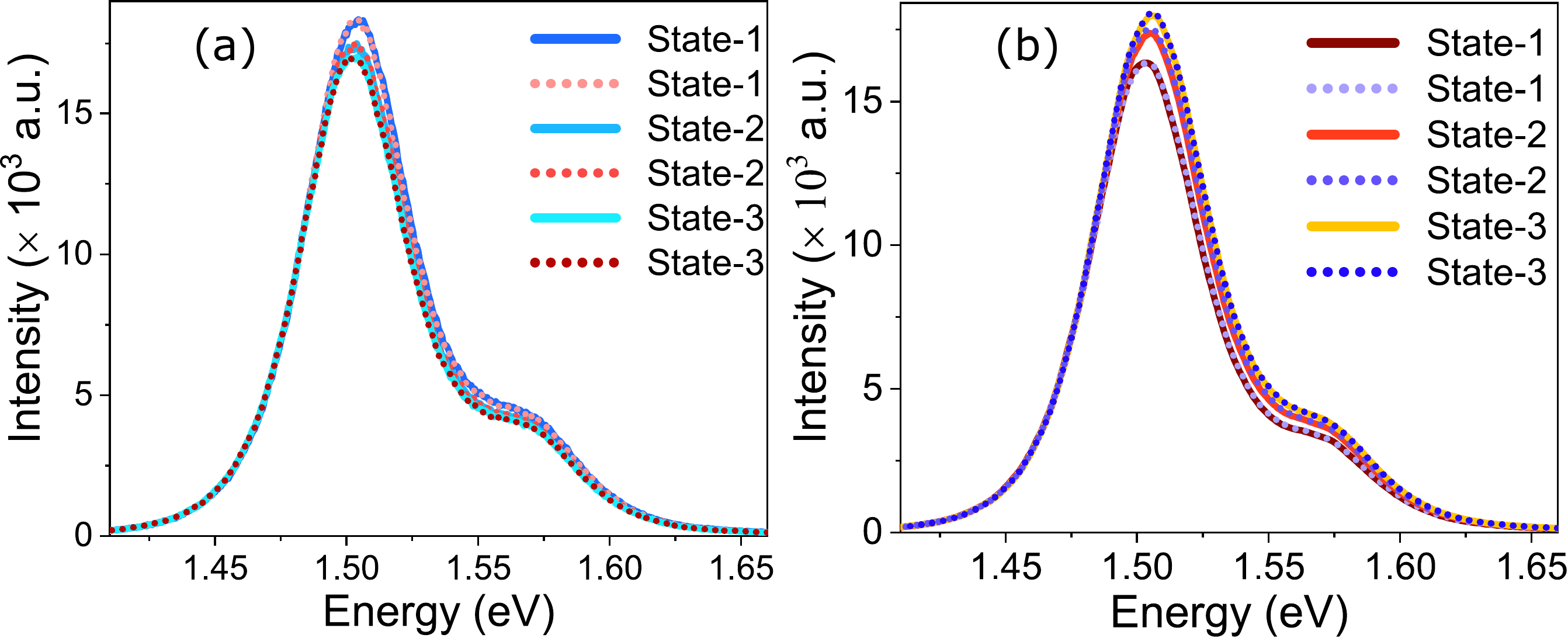}
			\caption{[supplement] PL spectra of state 1, state 2 and state 3 described in Fig. 5[main text]. The spectra corresponding to the starting of a minor loop are denoted by a solid line. The spectra corresponding to the ending of a minor loop are denoted by a dotted line. (a) and (b) correspond to the state described in Fig. 5[main text] and Fig. 6[supplement], respectively. The Pearson correlation coefficients are respectively 0.9998 (state 1, cooling), 0.99995 (state 1, heating), 0.9996 (state 2, cooling), 0.999988 (state 2, heating), 0.99995 (state 3, cooling), and 0.999991 (state 3, heating).}
		\end{center}
	\end{figure}

	\begin{figure}[!h]\label{FigSuppl:RPM_Difference}
		\begin{center}
			\includegraphics[scale=0.25]{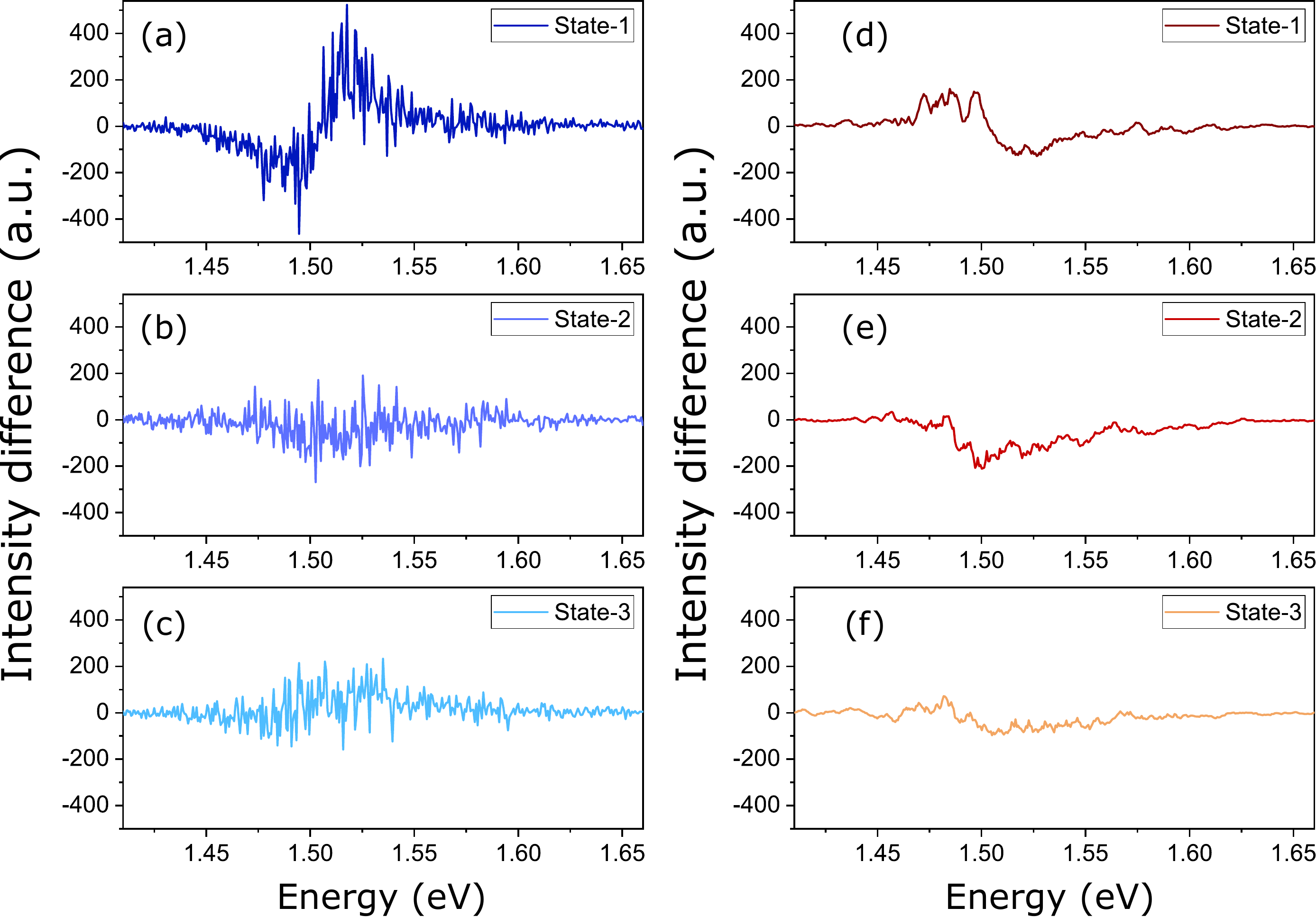}
			\caption{[supplement] (a)-(c) correspond (respectively for state 1, state 2, and state 3) to the difference between the two spectra shown in Fig. 7(a)[supplement]. (d)-(f) correspond (respectively for state 1, state 2, and state 3) to the difference between the two spectra shown in Fig. 7(b)[supplement].}
		\end{center}
	\end{figure}
	\begin{figure}[h!]
		\begin{center}
			\includegraphics[scale=0.3]{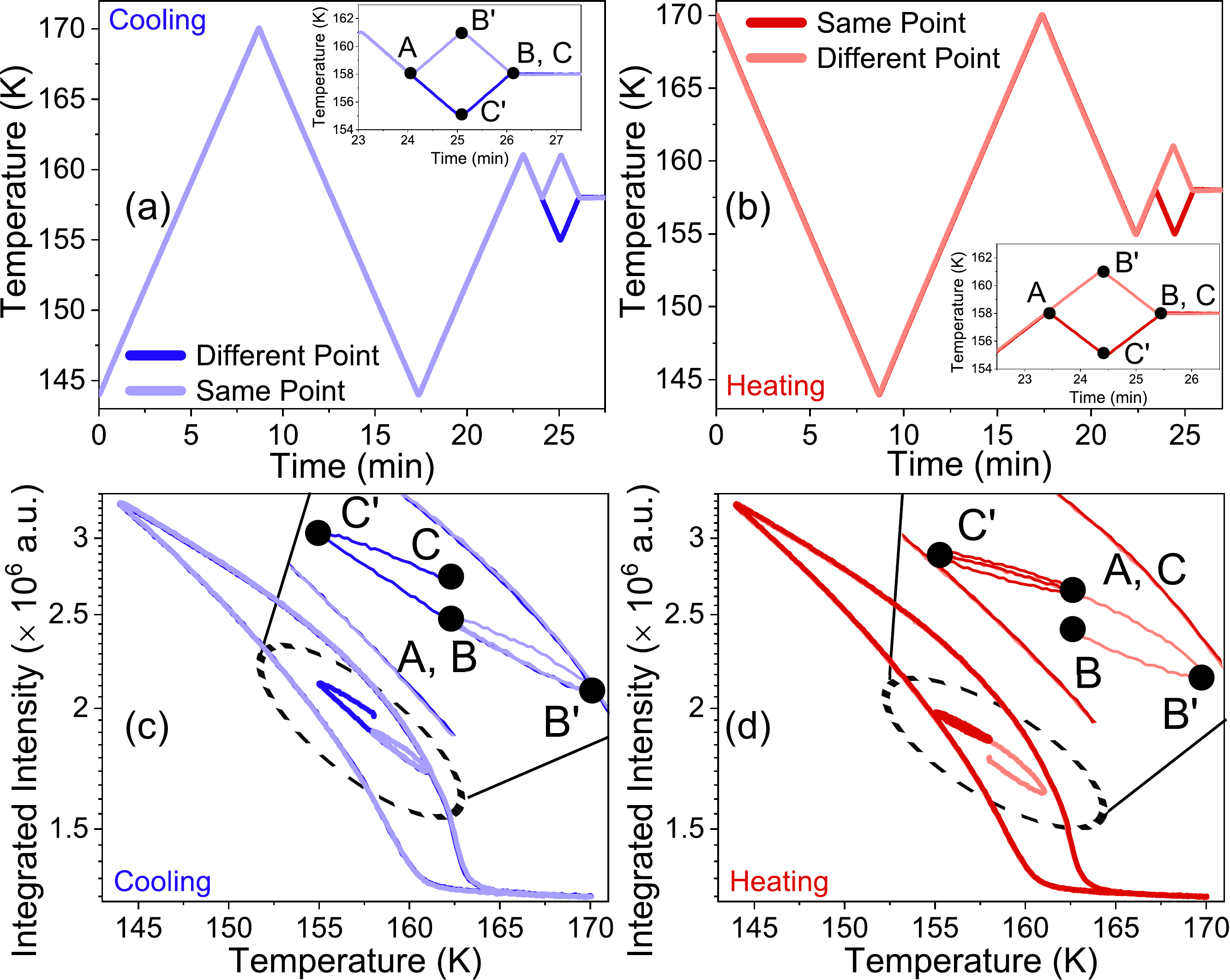}
			\caption{[supplement] This figure demonstrates how the observations of Fig. 3[main text] can be understood as a manifestation of the return-point memory. The temperature history used to create a metastable state A  at temperature 158 K by (a) a ``cooling protocol," and (b) a ``heating protocol".  We compare the effect of heating spike (AB$^{'}$B) and cooling spike (AC$^{'}$C) of 3 K for both cases. The corresponding integrated intensity profile is shown in (c) and (d). In (c) the system settles to different metastable states when encountering a cooling spike and in (d) when it encounters a heating spike. The system returns to the previous metastable state in the other two cases.}
		\end{center}
	\end{figure}
The actual PL spectra of state 1, state 2, and state 3 of Fig. 5[main text] and Fig. 6[supplement] are shown in Fig. 7[supplement]. The spectra corresponding to the starting of a minor loop are denoted by solid lines. The spectra corresponding to the ending of a minor loop are denoted by dotted lines. The difference between the spectra of the same state has been shown in Fig. 8[supplement].
	
Let us also try to describe the observed return-point memory behavior in another way that also directly reconciles the observations in Fig. 3[main text].
In Fig. 9[supplement], we prepare a metastable state called A at 158 K via two different thermal histories. Thermal history (i): 170 K $\rightarrow$ 144 K $\rightarrow$ 161 K $\rightarrow$ 158 K. This is described in Fig. 9(a)[supplement].  Thermal history (ii): 144 K $\rightarrow$ 170 K $\rightarrow$ 155 K $\rightarrow$ 158 K, in  Fig. 9(b)[supplement]. The state A with thermal history (i) is prepared twice and it is once given a $+$3 K thermal spike so that it evolves to a state A, following the path AB'B, and the next time given a $-$3 K thermal spike so that it evolves along AC'C---these are shown in Fig. 9(a)[supplement]. The state A with thermal history (ii) is also similarly prepared twice and it is once given a $+$3 K thermal spike so that it evolves to a state A, following the path AB'B and the next time given a $-$3 K thermal spike so that it evolves along AC'C---these are shown in Fig. 9(b)[supplement].
	
The difference in the integrated intensity of the PL spectra at A and B and A and C is shown in  Fig. 9(c)[supplement]. We find that for the thermal history in  Fig. 9(a)[supplement], a positive thermal spike (heating spike) does not cause an irreversible evolution of the system but a negative thermal spike (cooling spike) does. Since this metastable state is created by cooling (161 K $\rightarrow$ 158 K), any heating spike less than 3 K does not affect the system.  It can be seen in Fig. 9(c)[supplement] that the final state B after the heating spike coincides with state A but the final state C after the cooling spike is different from state A. Similarly, in the second metastable state Fig. 9(b)[supplement], negative thermal spike (cooling spike) does not perturb the system but positive thermal spike (heating spike) does. Since this metastable state is created by heating (155 K $\rightarrow$ 158 K), any cooling spike less than 3 K does not affect the system. In Fig. 9(d)[supplement], we show that the final state C after the cooling spike coincides with state A but the final state B after the heating spike is different from state A.
 \clearpage
 \subsection{Return-point memory: XRD}
 In Fig. 10[supplement], the return-point memory experiment is demonstrated 
for the spatially-averaged order parameter (phase fraction), both while the excursions out of the major hysteresis loop into the metastable phase are made while cooling [Figs. 10(a), 10(b)[supplement]] and heating [Figs. 10(c), 10(d)[supplement]]. 

\begin{figure}[!h]
		\begin{center}
			\includegraphics[scale=0.3]{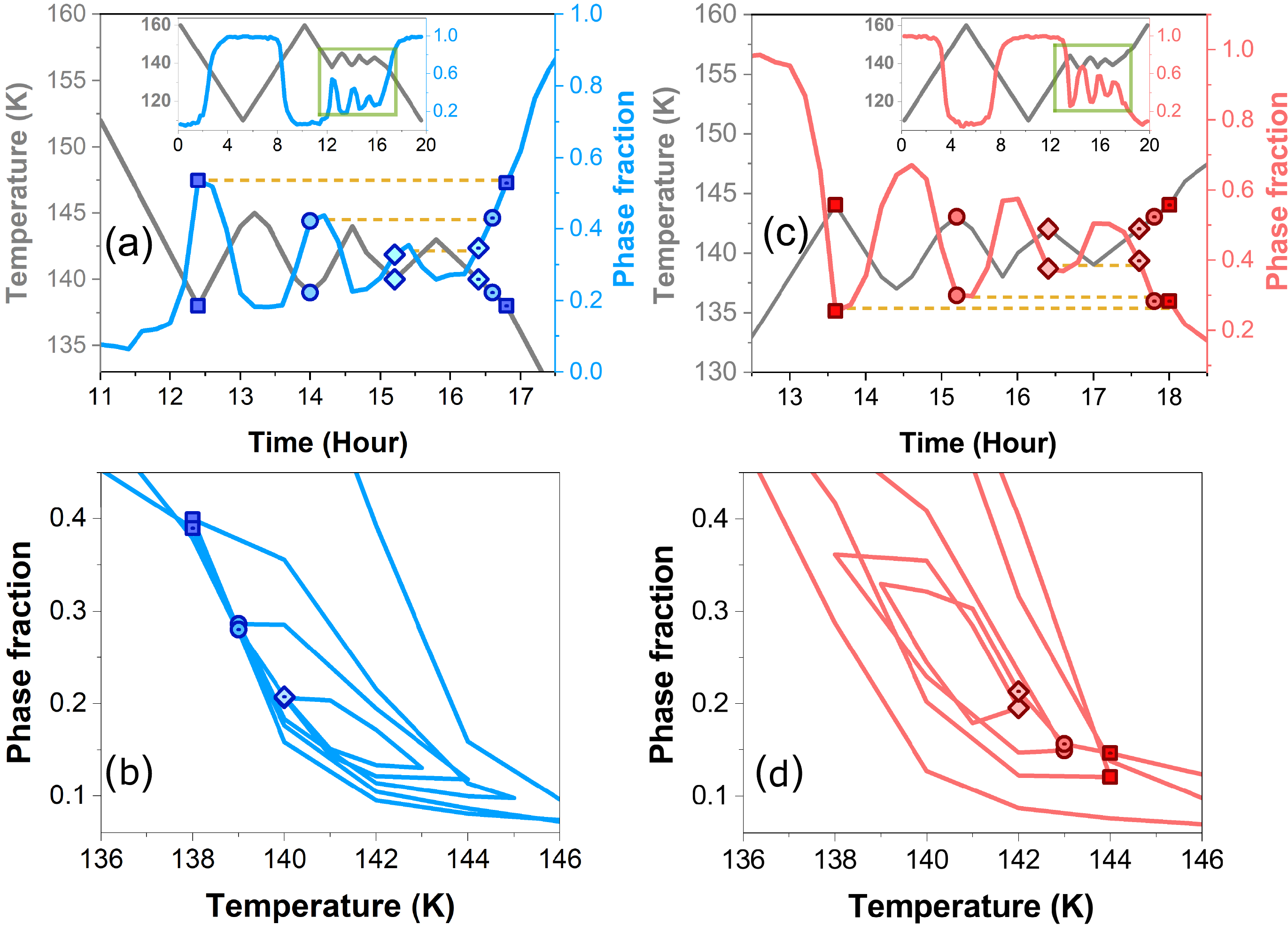}
			\caption{[supplement] Demonstration of return-point memory from XRD measurements. Panel (a) shows the thermal history (left axis) and the corresponding tetragonal-to-orthorhombic phase fraction (right axis) for the cooling part of the major loop. In panel (b) the data from panel (a) are now plotted in the temperature-phase fraction plane. Emphasis is placed on the minor loops, clearly marked in the corresponding insets. state 1, state 2, and state 3 are denoted by square, circle, and diamond symbols, respectively. Panels (c), and (d) show the corresponding measurements for the heating part of the major hysteresis loop. Note that the transition temperature for this sample is slightly different from the sample used for PL measurements.}
		\end{center}
	\end{figure}

 \clearpage
\section{Preisach Model: Hysteron microstructure and the return-point memory effect}
The Preisach model is a phenomenological but quantitative model that well-captures the return-point memory and other features of the metastable hysteretic phase \cite{Mayergoyz_PRL1986}. The Preisach model (and its many variants and extensions) [e.g. Chapter 3 in Volume 1 of Ref. \onlinecite{Ortin_Planes_Delaey}] has been extensively used to describe the magnetic field-dependent hysteresis in ferromagnets and also the metastable phase in systems such as shape-memory alloys [e.g. Chapter 5 in Volume III of Ref. \onlinecite{Ortin_Planes_Delaey}] and transition metal oxides which show thermal hysteresis around the first order thermodynamic phase transition. 

It is emphasized that the model is based on an ``engineering approach" in that it quantitatively describes the behavior of the metastable phase but has no clear microscopic statistical mechanical basis. Nevertheless, the Preisach model serves as a very powerful modeling and visualization tool that captures much of the intriguing phenomenology of the metastable phase. 
\begin{figure}[h!]
		\begin{center}
			\includegraphics[scale=0.3]{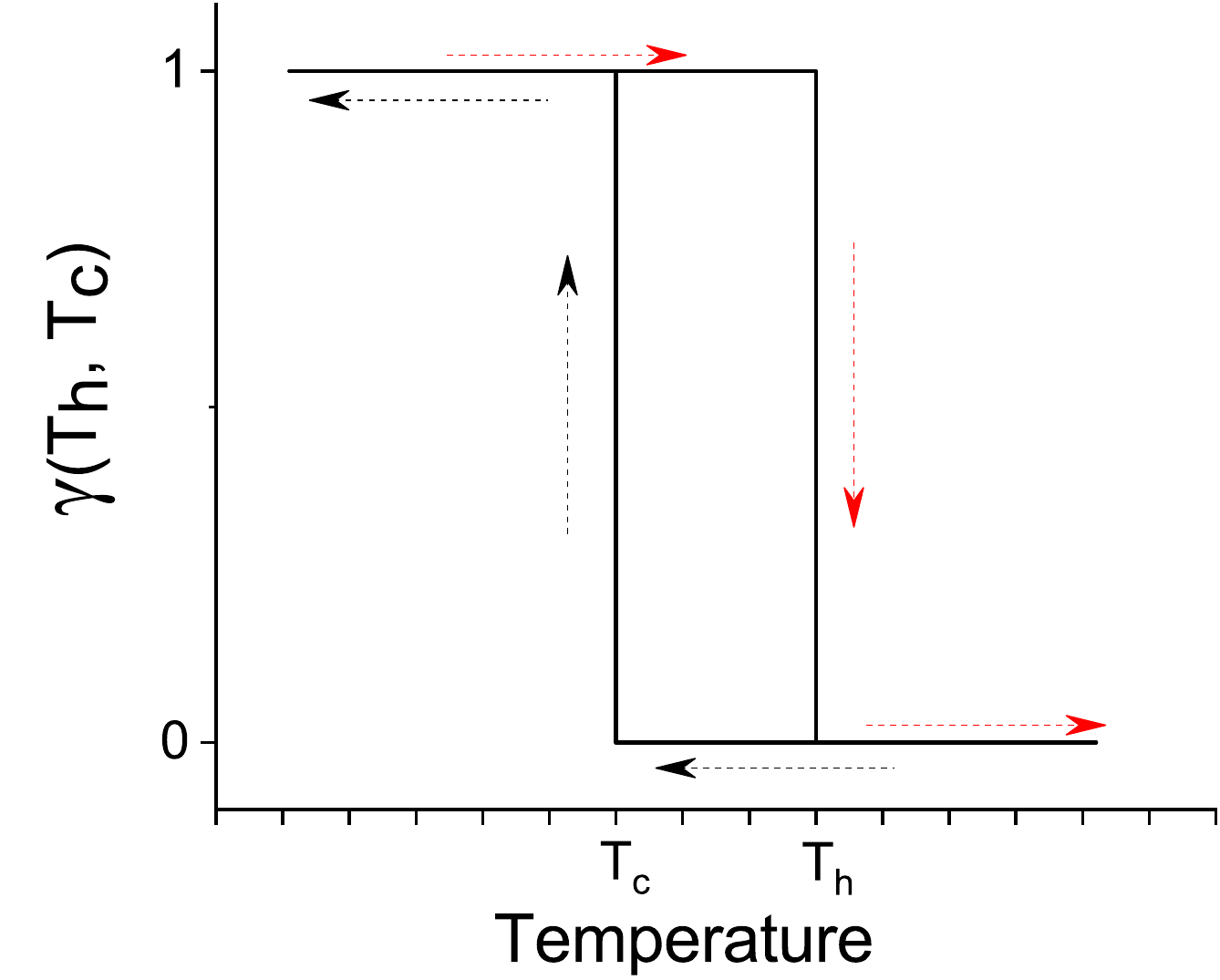}
			\caption{[supplement] An elementary hysteron $\gamma (T_c, T_h)$ characterized by the two switching temperatures $T_c$ and $T_h$ corresponding to the transition while cooling and heating respectively. The metastable phase is constructed as an ensemble of such hysterons.}
		\end{center}
	\end{figure}
In the Preisach model, the metastable phase---where the observables are history-dependent and multivalued---is visualized as a sum of noninteracting ``hysterons". A hysteron is a bistable switching operator, capable of taking two values, say, 0 (off) and 1 (on). Furthermore, each hysteron is characterized by two numbers, the values at which the switching occurs as the parameter causing the switch is changed in the forward and the reverse direction, respectively.  For the case of thermal hysteresis, we represent a hysteron as $\gamma(T_h, T_c)$, $T_h$, and $T_c$ being the switching temperatures for that specific hysteron while heating and cooling, respectively. Such an elementary hysteron $\gamma (T_h, T_c)$ is depicted in Fig. 11[supplement].

Consider the experiment where we monitor the integrated PL intensity $I$ to characterize the hysteretic phase transition in MAPbI$_3$. Using the Preisach model, one may describe (up to a scale factor) the history-dependent integrated PL intensity $I$ as a weighted contribution from an ensemble of hysterons, viz., 
\begin{equation}\label{Eqn. Preisach}
I = \sum_{T{_h} \geq T{_c}} \mu(T{_h}, T{_c}) \times \gamma(T{_h}, T{_c}).
\end{equation}
Here the Preisach measure $\mu(T{_h}, T{_c})$ represents the sample-specific hysteron density that must be experimentally determined. Since $T_h\geq T_c$, Eq, \ref{Eqn. Preisach}[supplement] denotes a triangle in a plane with $T_c$ and $T_h$ forming the orthogonal axes. 

In practice, $\mu(T{_h}, T{_c})$ is estimated by measuring the first-order reversal curves [e.g. Chapter 3 in Volume 1 of Ref. \onlinecite{Ortin_Planes_Delaey}]. Figure 12[supplement] summarizes such a measurement for our MAPbI$_3$ sample. In Fig. 12(a)[supplement], the multivalued PL intensity is plotted as the function of temperature. Starting with the temperature of 144 K, the sample is heated to a reversal temperature $T_R$ and brought back to the low-temperature orthorhombic phase at 144 K. In this way, the temperature is cycled many times with the reversal temperature $T_R$ progressively lowered in each cycle. For $T_R\lesssim 165 $ K, the phase transformation on reaching $T_R$ is only partial and the fraction of the transformed material is a function of both the sample temperature $T$ and the path followed to reach $T$ (that is characterized by $T_R$). One can see that such reversal curves can yield the Preisach density function $\mu(T{_h}, T{_c})$
can be determined by taking the second derivative 
$$\mu(T{_h}, T{_c})=\pm\frac{\partial^2 I}{\partial T_c\partial T_h}.$$ The positive and negative signs correspond to the heating and the cooling reversal curves respectively. 
\begin{figure}[!t]
		\begin{center}
			\includegraphics[scale=0.2]{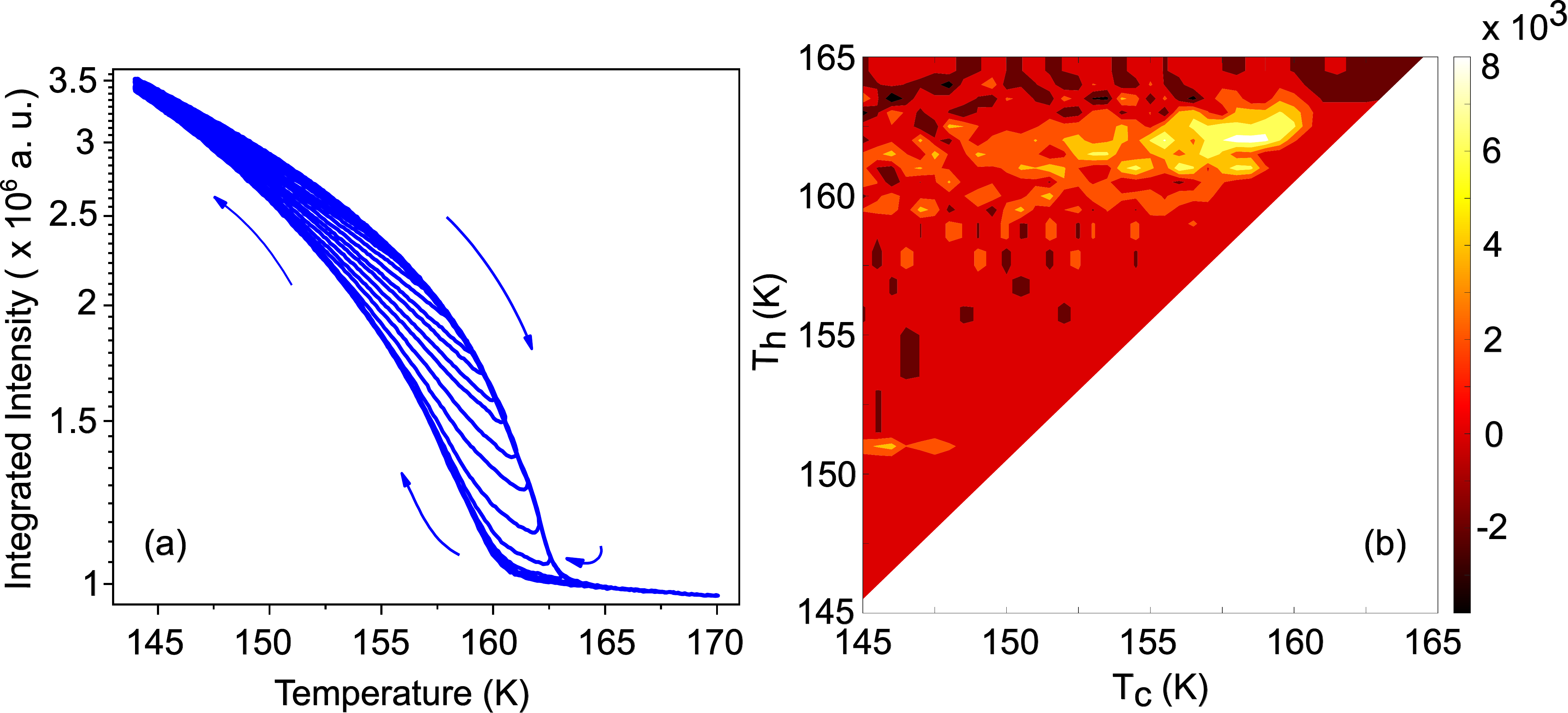}
			\caption{[supplement] Preisach Model of Hysteresis. (a) The method of first-order reversal curve (FORC) for cooling is used to calculate the value of the Preisach function. After completing the first cycle, to create first-order reversal curves, the low-temperature end of the temperature cycle is fixed at 144 K, and the value of the other high-temperature end is progressively decreased from 165 K to 151 K creating minor loops within the region of metastability. (b) Shows a picture of a hysteron that switches between states at different temperatures for heating (at $T_h$) and cooling (at $T_c$). And the colored triangle shows the contour plot of the Preisach function. Here each point depicts a hysteron with corresponding coordinates as hysteron’s transition temperatures ($T_h$ and $T_c$) such that $T_h$ $\geq$ $T_c$ for all hysterons and it is accompanied with only hysteretic portion of limiting cycle which lies between 145 K to 165 K. The peak value of the Preisach function signifies the maximum contribution of the hysterons corresponding to the region of metastability with an abrupt change in intensity value.}
		\end{center}
	\end{figure}
Figure 12(b)[supplement] shows the Preisach function $\mu(T{_h}, T{_c})$ approximated from the reversal curves in Fig. 12(a)[supplement]. In this contour plot, each point depicts a hysteron with coordinates ($T_c, T_h$), and the fractional weight of the hysteron to the total integrated intensity is given by $\mu(T{_h}, T{_c})$. Since $T_h\geq T_c$, we have a right-angle triangle. 

Excursions made by partial/minor hysteresis loops correspond to selective switching of a few hysterons whereas the traversal from low to high temperatures (or vice versa) along the major hysteresis loop switches all the hystereons of the triangle. The peak in the contour plot [Fig. 12(b)[supplement]] corresponds to the highest density of hysterons. 
\begin{figure*}[h!]
		\begin{center}
			\includegraphics[scale=0.13]{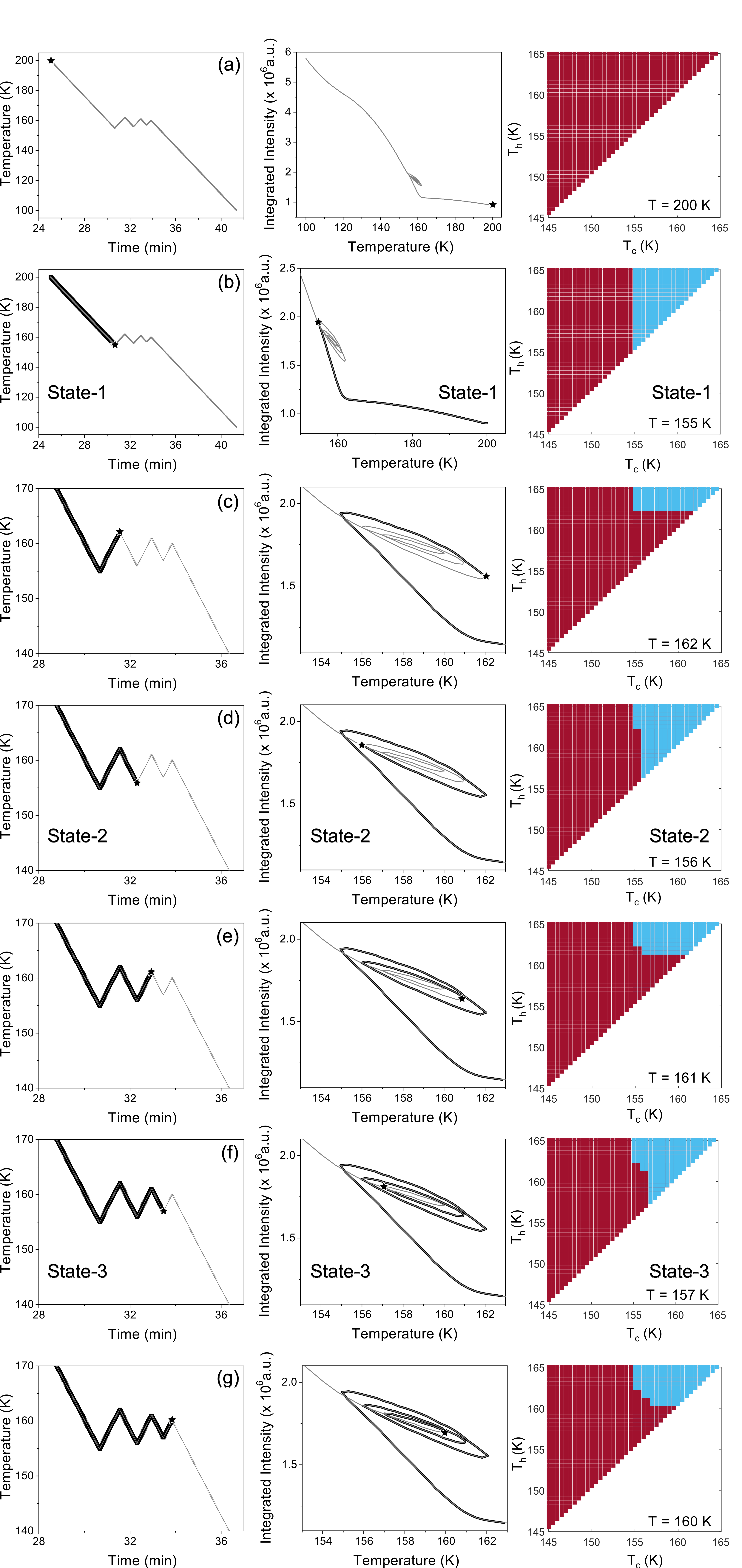}
			\caption{[supplement] Visualizing the return-point memory behavior. The figure shows the interpretation of the return-point memory experiment [Fig. 5 (main text)] as the evolution of the Preisach triangle comprising of elementary hysterons. (a)-(g) The sequence of nested thermal cycles hierarchically encodes a state and this encoding is sequentially erased in the sequence shown in (g)-(k) on the next page. The 1$^{st}$, 2$^{nd}$, and 3$^{rd}$ column of each row corresponds to the temperature variation, the accompanying intensity evolution, and the microstate in the Preisach triangle, respectively. The current state of the system is represented by an asterisk (\boldsymbol{$\star$}). In the first two columns, the road map of the complete experiment is depicted as a thin line and the sequence traversed up to the particular point is depicted as a thick line.  (a) At $T$ = 200 K, the system is in the tetragonal phase and all hysterons have an OFF state. The Preisach triangle is completely brown. (b) As temperature varies from 200 K to 155 K a blue region sweeps the Preisach triangle up to 155 K and the hysterons in the region 200 K-150 K are switched ON. We have reached state 1 of Fig. 5[main text]. (c) As the temperature sweep reverses a horizontal line sweeps the Preisach plane from $T_h$ = 155 K to $T_h$ = 162 K. All the previously ON hysterons are turned OFF (and OFF stay OFF).  In this way, as temperature varies from 162 K to 156 K to 161 K to 157 K to 160 K, we reach state 2 and state 3 (of Fig. 5[main text]) in panels (d) and (f), respectively. Note that the scale in the first two columns of (a), (b), and the rest of the rows is different. The figure continues below.}
		\end{center}
	\end{figure*}

\setcounter{figure}{12}
\begin{figure*}[h!]
		\begin{center}
			\includegraphics[scale=0.15]{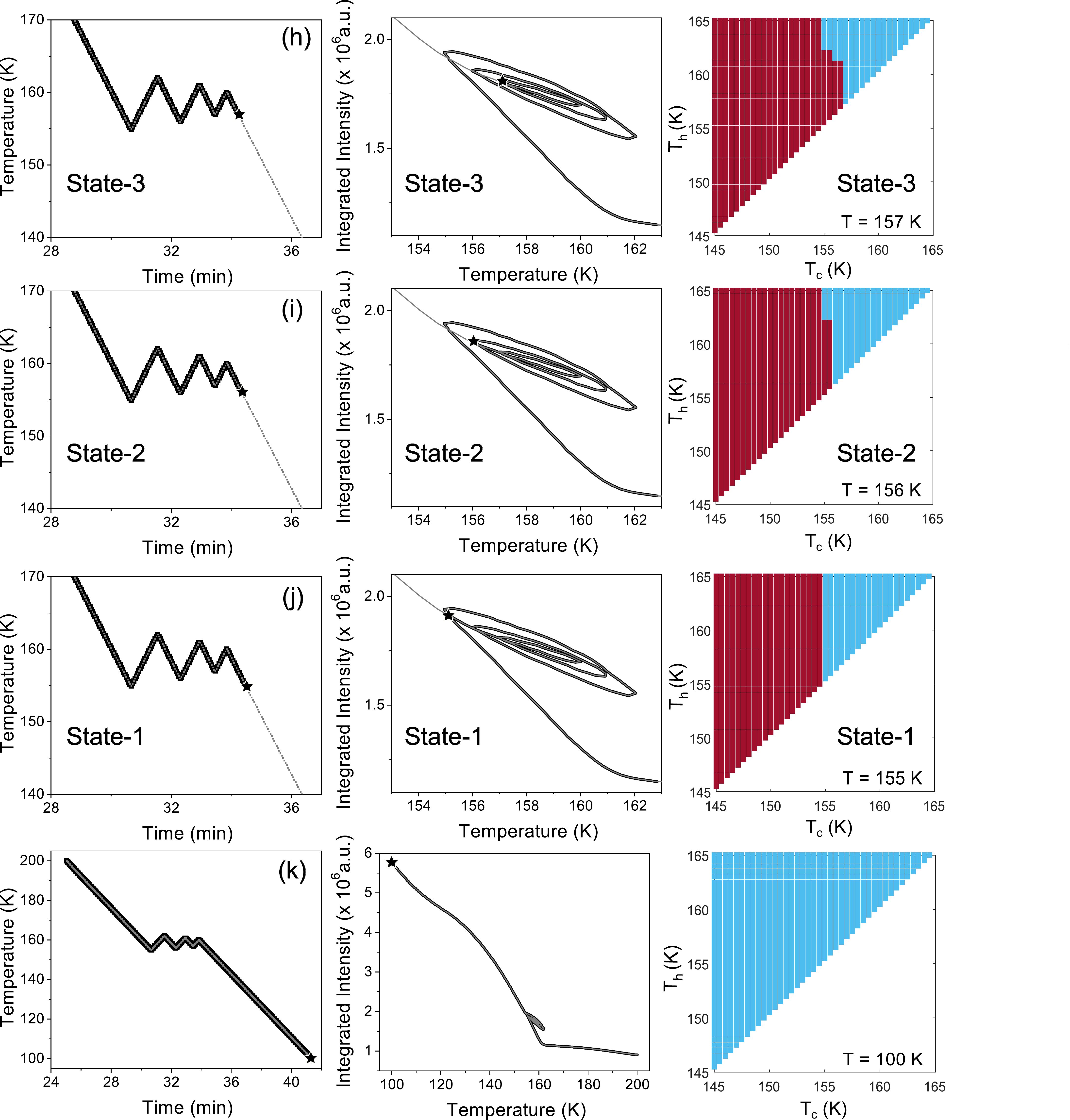}
			\caption{[supplement] (continued from the previous page)  Visualizing the return-point memory behavior. The figure shows the interpretation of the return-point memory experiment [Fig. 5 (main text)] as the evolution of the Preisach triangle comprising elementary hysterons. On the previous page (a)-(g) showed the hierarchical encoding of state 1, state 2, and state 3 of Fig. 5[main text]. This memory is sequentially erased in the sequence shown in (g)-(k). Finally, as the temperature is lowered to 200 K all the hysterons get to the ON state. Note the difference in scale between figures in the first two columns of (k) and the rest of the rows.}
		\end{center}
	\end{figure*}
We now demonstrate in Fig. 13[supplement] how the sequence of the return-point memory experiment (Fig. 5[main text]) is uniquely represented in the Preisach triangle. The brown color at a coordinate ($T_c, T_h$) represents a hysteron $\gamma(T_h, T_c)$ in the high-temperature tetragonal phase whereas the blue color represents the low-temperature orthorhombic phase. Thus when the sample is completely in the low-temperature orthorhombic phase ($T<145$ K), we have a filled blue triangle and when it is in a single-phase tetragonal state ($T>165$ K), we have a fully brown triangle.    

Figure 13[supplement] has 11 rows, labeled (a) through (k), depicting the state of the system at a given temperature during a particular stage of the return-point memory experiment described in Fig. 5[main text]. 
The  1$^{st}$, 2$^{nd}$, and 3$^{rd}$ column of each row shows the temperature, PL intensity, and the Preisach triangle, respectively. Specifically, in the first column of each row, the thermal path to be traversed through the entire experiment is shown by a thin line and a superposed thick line represents the path traversed up to that particular panel. The current state is marked with an asterisk (\boldsymbol{$\star$}). 

The second column of Fig. 13[supplement], showing the PL intensity, similarly represents the variation of the PL intensity as a thin line and the path traversed so far as a superposed thicker line. The present state of the system is represented as \boldsymbol{$\star$}. Note the difference in scale in the data shown in the middle column of (a), (b), and (k) rows and the rest of the rows. 

The initial state prepared in  Fig. 13(a)[supplement] corresponds to a pure tetragonal phase at 200 K and thus all the hysterons are OFF and the triangle is completely brown. 
In general, one may start with any configuration with a certain fraction of hysterons in the ON state. The hysterons in the Preisach plane respond to the temperature change in the following way. Lowering the temperature from, say, $T_i$ to $T_f$ leads to a vertical line sweeping the Preisach plane right to left from $T_i$ to $T_f$.  All the hysterons in the swept region go to the ON state. Hysterons which were already ON, stay ON. 

Raising the temperature from, say, $T_i$ to $T_f$ on the other hand leads to a horizontal line sweeping the Preisach plane upwards from $T_i$ to $T_f$.  All the hysterons in this region go to the OFF state. Hysterons which were already OFF, stay OFF. 

This algorithm allows us to store any metastable state as a unique two-color pattern in the Preisach triangle. The hierarchy is encoded in the number of right-angle edges inside the triangle. 

In Fig. 13[supplement], the return-point memory (Fig. 5[main text]) is established in the hysteron picture since the triangles in (b) and (j) (which both depict state 1), the triangles in (d) and (i) (state 2), and the triangles in (f) and (h) (state 3)  are identical.  
\newpage
\section{Number of accessible metastable states}
Based on the above observations, it is evident that many points in the metastable state can be accessed by suitably manipulating the thermal history. Furthermore, the analysis based on hysterons in the Preisach triangle allows for a unique description of such a metastable state. Given that these states are athermal (viz., they don't diffuse away over time) and obey return-point memory suggests a hierarchical memory \cite{Perkovic_Sethna}---different from the thermal-breach memory that we have discussed. 

In the following, we tentatively discuss the number of states in such a memory in our material. This number may be estimated from the number of possible nested temperature reversals within the hysteresis window, under the constraints set by temperature stability and the degree of the athermal character of the material. 

Thus the hysteresis window $\Delta T$ may be broken into $N$ temperature intervals of size $\delta T$, such that $\Delta T=N\delta T$. We will thus have a Preisach triangle with $N^2/2$ hysterons. This temperature resolution $\delta T$ is constrained by the measurement precision, the temperature stability, and the stability of the material against isothermal evolution. 

For the MAPbI$_3$ sample used for PL measurements, the metastable phase roughly extends between $144-164$ K. Thus $\Delta T\approx 20$ K. Further assuming $\delta T\approx  1$ K, we have $N\approx 20$. 

Given that the various metastable states are accessible via temperature reversals and the return-point memory property holds, it is not hard to show that the number of states ($\#_{\Delta T, \delta T}$) is\footnote
{
To explicitly verify this formula, assume a much smaller $\Delta T$ = 150 K - 146 K = 4 K; $\delta T=1$ K.  
Consider the situation where one starts from (or above) 150 K and the first temperature change involves cooling. The sample may be cooled by 1 K ($p=1$), 2 K ($p=2$), 3 K ($p=3$) or 4 K ($p=4$) below 150 K. Let us explicitly count the states for each case. 

When the first cooling step before reversal is $1$ K, viz, $p=1$ term in the above summation, we have
\[
[p=1, 2^0\,\textrm{states} ]\rightarrow \begin{pmatrix}
    150 \rightarrow 149\, \textrm{K}
        \end{pmatrix}.
\]
Starting from 150 K, when the first cooling step before reversal is $2$ K, viz, $p=2$ term in the above summation, we have
\[
[p=2, 2^1\,\textrm{states} ]\rightarrow \begin{pmatrix}
    150 \rightarrow 148\, \textrm{K}\\
    150 \rightarrow 148 \rightarrow 149 \, \textrm{K}
        \end{pmatrix}.
\]
Starting from 150 K, when the first cooling step before reversal is $3$ K, viz, $p=3$ term in the above summation, we have

\[
[p=3, 2^2\,\textrm{states} ]\rightarrow \begin{pmatrix}
        150 \rightarrow 147 \,\textrm{K}\\
\begin{pmatrix}150 \rightarrow 147 \rightarrow 148 \,\textrm{K}\\ 150 \rightarrow 147 \rightarrow 149 \,\textrm{K}\end{pmatrix}\\ 150 \rightarrow 147 \rightarrow 149\rightarrow 148\,\textrm{K}
        \end{pmatrix}.
\]
Starting from 150 K, when the first cooling step before reversal is $4$ K, viz, $p=4$ term in the above summation, we have

\[
[p=4, 2^3\,\textrm{states} ]\rightarrow \begin{pmatrix}
        150 \rightarrow 146 \,\textrm{K}\\
\begin{pmatrix}150 \rightarrow 146 \rightarrow 147 \,\textrm{K}\\ 150 \rightarrow 146 \rightarrow 148 \,\textrm{K} \\ 150 \rightarrow 146 \rightarrow 149 \,\textrm{K}\end{pmatrix}\\ 
\begin{pmatrix}
150 \rightarrow 146 \rightarrow 148 \rightarrow 147\,\textrm{K}\\
150 \rightarrow 146 \rightarrow 149 \rightarrow 147\,\textrm{K}\\
150 \rightarrow 146 \rightarrow 149 \rightarrow 148\,\textrm{K}
\end{pmatrix}\\

150 \rightarrow 146 \rightarrow 149\rightarrow 147 \rightarrow 148\,\textrm{K}
        \end{pmatrix}.
\]
The states with the same number of temperature reversals ($\rightarrow$) have been bracketed together to highlight the correspondence with the coefficients of the binomial expansion; the multiplicity of the states for a given $p$ corresponds to the $p^\text{th}$ row of the Pascal's triangle. 
Thus for $N=4$ ($\Delta T=4$ K and $\delta T=1$ K), we have $\#_{4 K,1K}=2^0+2^1+2^2+2^3=15=2^4-1$, in agreement with Eq. \ref{Eqn:statecount}.
}
\begin{equation}\label{Eqn:statecount}
\#_{\Delta T, \delta T}=\sum_{p=1}^N 2^{p-1}=2^N-1; \;\;\text{where}\; N=\Delta T/\delta T.
\end{equation}
Thus a small macroscopic volume of the sample can have $2^{20}-1\approx 10^{6}$ memory states.  

Note that these hysterons are not independently addressable bits of conventional memory but, as we have seen in Fig. 13[supplement], combine in a complex and irreducible manner to encode a hierarchical structure. Since this is conceptually very different from the usual artificial memories, it is not quite clear how such an organized structure may be useful, if at all. Also there is no simple way of reading this memory without destroying it. 

\newpage
\section{Logarithmic relaxation}
We have seen in Figs. 2(b), 2(e)[main text] that the spectral shape in the metastable phase shows a very high degree of reproducibility over time, as measured by PCC, if the temperature is held constant.  

The PL intensity, on the other hand, shows a small change $\delta I(t)$ with time. This is shown in Figs. 14(a), 14(c)[supplement]. Figure 14(a)[supplement] corresponds to the intensity change for the experiments in Figs. 2(a)-2(c)[main text] and Fig. 14(c)[supplement] corresponds to the intensity change for the experiments in Figs. 2(d)-2(f)[main text]. We further find that the intensity change fits the function $\delta I(t)=S\cdot\ln{(1+t/t_0)}$, where $S$ and $t_0$ are free parameters, extremely well [Figs. 14(b), 14(d)[supplement]]. 

 \begin{figure}[h!]
		\begin{center}
			\includegraphics[scale=0.19]{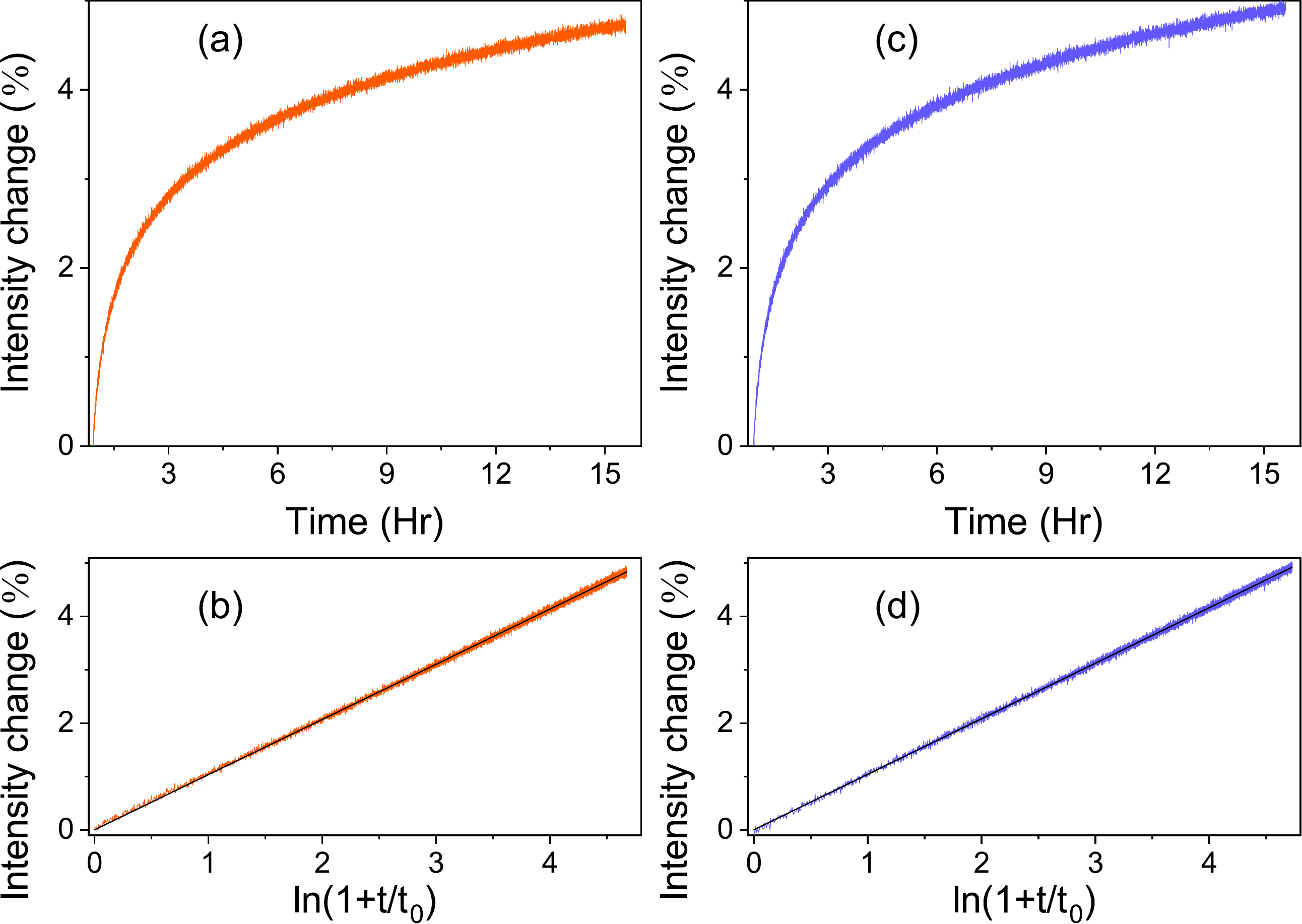}
			\caption{[supplement] In the metastable phase under isothermal conditions, a few percent monotonic increase in PL intensity over time is observed. (a),(b) Show the temporal evolution of the PL intensity for the experiment in Figs. 2(a)-2(c)[main text]. (c),(d) The same for Figs. 2(d)-2(f)[main text]. (b),(d) The intensity change shows a remarkably good fit to the function $\delta I(t)=S\cdot\ln{(1+t/t_0)}$.}
		\end{center}
	\end{figure}

We emphasize that this temporal evolution of the PL intensity is unrelated to phase ordering. This can be clearly argued from the fact that in both Figs. 14(a), 14(c)[supplement], we observe an increase in intensity over time, although the states are prepared with contrary thermal protocols. If the PL intensity change were reflecting the isothermal phase evolution, we should have observed a reduced intensity over time in Figs. 14(a), 14(b)[supplement]  and increased intensity over time in Figs. 14(c), 14(d)[supplement], as would be expected from Fig. 2(a)[main text] and Fig. 2(d)[main text], respectively. The monotonic increase of the PL intensity with time in both cases perhaps reflects the mechanical accommodation associated with the structural transition and the healing of defects. This is unrelated to the phase evolution as the spectral shape is arrested to a much higher degree.

This logarithmic relaxation (``creep") is one of the hallmarks of nonlinear complex behavior that has been seen in a variety of contexts from magnetic viscosity \cite{Muller_Viscosity}, flux decay in superconductors \cite{Yeshurun_RMP} to crumpled paper \cite{Balankin_PRE2011}, spin glasses and electron glasses \cite{Eisenbach_ElectronGlass}, ultimately going back to Boltzmann's elastic after-effect \cite{Rankine_1907}. 

Here we have simply reported on this interesting phenomenon for completeness. A detailed discussion, on the other hand, would be quite outside the theme of the present work.


\begin{thebibliography}{99}
\bibitem{Chaikin}P. M. Chaikin and T. C. Lubensky, \textcolor{blue}{\href{https://doi.org/10.1017/CBO9780511813467}{Principles of Condensed Matter Physics}} (Cambridge University Press, Cambridge, England, 1995).

\bibitem{Stern-Murugan}M. Stern and A. Murugan, Learning without neurons in physical systems, \textcolor{blue}{\href{https://www.annualreviews.org/doi/10.1146/annurev-conmatphys-040821-113439}{Annu. Rev. Condens. Matter Phys. {\bf 14}, 417 (2023)}.}	

\bibitem{Keim_RMP}N. C. Keim, J. D. Paulsen, Z. Zeravcic, S. Sastry, and S. R. Nagel, Memory formation in matter, \textcolor{blue}{\href{https://journals.aps.org/rmp/abstract/10.1103/RevModPhys.91.035002}{Rev. Mod. Phys. {\bf 91}, 035002 (2019)}.}

\bibitem{Bensea_PNAS}H. Bensea and M. van Hecke, Complex pathways and memory in compressed corrugated sheets, \textcolor{blue}{\href{https://doi.org/10.1073/pnas.2111436118}{Proc. Nat. Acad. Sci. U.S.A. {\bf 118}, e2111436118 (2021).}}	

\bibitem{Mungan_PNASReview}M. Mungan, Putting memories on paper,  \textcolor{blue}{\href{https://doi.org/10.1073/pnas.2208743119}{Proc. Nat. Acad. Sci. U.S.A., {\bf 119}, e2208743119 (2022)}}.	

\bibitem{Pershin-DiVentra}Y. V. Pershin and M. Di Ventra, Memory effects in complex materials and nanoscale systems, \textcolor{blue}{\href{http://dx.doi.org/10.1080/00018732.2010.544961}{Adv. Phys. {\bf 60}, 145 (2011)}}.

\bibitem{Nova_SrTio3}T. F. Nova, A. S. Disa, M. Fechner, and A. Cavalleri, Metastable ferroelectricity in optically strained SrTiO$_3$, \textcolor{blue}{\href{https://www.science.org/doi/10.1126/science.aaw4911}{Science {\bf 364}, 1075 (2019)}}.

\bibitem{Ordonez-Miranda}J. Ordonez-Miranda, Y. Ezzahri, J. A. Tiburcio-Moreno, K. Joulain, and J. Drevillon, Radiative thermal memristor,  \textcolor{blue}{\href{https://journals.aps.org/prl/abstract/10.1103/PhysRevLett.123.025901}{Phys. Rev. Lett. {\bf 123}, 025901 (2019)}}.

\bibitem{Jiang_FerroelectricPRL2023}Y. Jiang, X. Ma, L. Wang, J. Zhang, Z. Wang, R. Zhao, G. Liu, Y. Li, C. Zhang, C. Ma, Y. Qi, L. Wu, and J. Gao, Observation of electric hysteresis, polarization oscillation, and pyroelectricity in nonferroelectric p--n heterojunctions, \textcolor{blue}{\href{https://doi.org/10.1103/PhysRevLett.130.196801}{Phys. Rev. Lett. {\bf 130}, 196801 (2023)}}.

\bibitem{Barker_ProcRoySoc1983}J. A. Barker, D. E. Schreiber, B. G. Huth, and D. H. Everett, Magnetic hysteresis and minor loops: Models and experiments, \textcolor{blue}{\href{http://dx.doi.org/10.1098/rspa.1983.0035}{Proc. R. Soc. A {\bf 386}, 251 (1983)}}.

\bibitem{Mayergoyz_2003Book}I. Mayergoyz, {\it Mathematical Models of Hysteresis and their Applications} 2nd ed. (Elsevier, New York 2003).

\bibitem{Bertotti}G. Bertotti, \textcolor{blue}{\href{https://www.sciencedirect.com/book/9780120932702/hysteresis-in-magnetism}{Hysteresis in Magnetism}} (Academic Press, San Diego 1998).

\bibitem{Kronmuller}H. Kronm\"uller and M. F\"ahnle, \textcolor{blue}{\href{https://www.cambridge.org/in/universitypress/subjects/physics/condensed-matter-physics-nanoscience-and-mesoscopic-physics/micromagnetism-and-microstructure-ferromagnetic-solids?format=PB&isbn=9780521120470}{Micromagnetism and the Microstructure of Ferromagnetic Solids}} (Cambridge University Press,  Cambridge, England, 2003).

\bibitem{Reichhardt_2023} C. Reichhardt, I. Regev, K. Dahmen, S. Okuma, and C. J. O. Reichhardt, Reversible to irreversible transitions in periodic driven many-body systems and future directions for classical and quantum systems, \textcolor{blue}{\href{https://doi.org/10.1103/PhysRevResearch.5.021001}{Phys. Rev. Res. {\bf 5}, 021001 (2023)}}.

\bibitem{Candela2023_GranularMemory} D. Candela, Complex memory formation in frictional granular media, \textcolor{blue}{\href{https://doi.org/10.1103/PhysRevLett.130.268202}{Phys. Rev. Lett. {\bf 130}, 268202 (2023)}}.

\bibitem{Brenner_MAPbI3}T. M. Brenner, D. A. Egger, L. Kronik, G. Hodes, and D. Cahen, Hybrid organic--inorganic perovskites: Low-cost semiconductors with intriguing charge-transport properties,  \textcolor{blue}{\href{https://www.nature.com/articles/natrevmats20157}{Nat. Rev. Mater. {\bf 1}, 15007 (2016)}}.

\bibitem{Hutter_MaPbI3} M. Hutter, M. C. G\'elvez-Rueda, A. Osherov, V. Bulovic, F. C. Grozema, S. D. Stranks, and T. J. Savenije, Direct-indirect character of the bandgap in methylammonium lead iodide perovskite, \textcolor{blue}{\href{https://www.nature.com/articles/nmat4765}{Nat. Mat. {\bf 16}, 115 (2017)}}.

\bibitem{MAPBI3_Structural}M. Keshavarz, M. Ottesen, S. Wiedmann, M. Wharmby, R. K\"uchler, H. Yuan, E. Debroye, J. A. Steele, J. Martens, N. E. Hussey, M. Bremholm, M. B. J. Roeffaers, and J. Hofkens, Tracking structural phase transitions in lead-halide perovskites by means of thermal expansion, \textcolor{blue}{\href{https://onlinelibrary.wiley.com/doi/full/10.1002/adma.201900521}{Adv. Mater. {\bf 31}, 1900521 (2019)}}.

\bibitem{Abdi-Jalebi_MAPbI3}M. Abdi-Jalebi, Z. Andaji-Garmaroudi, S. Cacovich,  \textit{et al.}, Maximizing and stabilizing luminescence from halide perovskites with potassium passivation,  \textcolor{blue}{\href{https://doi.org/10.1038/nature25989}{Nature (London) {\bf 555}, 497 (2018)}}.

\bibitem{Akkerman}Q. A. Akkerman, G. Rain\'o, M. V. Kovalenko, and L. Manna, Genesis, Challenges and opportunities for colloidal lead halide perovskite nanocrystals,  \textcolor{blue}{\href{https://www.nature.com/articles/s41563-018-0018-4}{Nat. Mat. {\bf 17}, 394 (2018)}}.

\bibitem{Sharada}G. Sharada, P. Mahale, B. P. Kore, S. Mukherjee, M. S. Pavan, C. De, S. Ghara, A. Sundaresan, A. Pandey, T. N. Guru Row, and D. D. Sarma, Is CH$_3$NH$_3$PbI$_3$ polar?,  \textcolor{blue}{\href{https://pubs.acs.org/doi/full/10.1021/acs.jpclett.6b00803}{ J. Phys. Chem. Lett.  {\bf 7}, 2412 (2016)}}.

\bibitem{Ashutosh_2019}A. Mohanty, D. Swain, S. Govinda, T. N. Guru Row, and D. D. Sarma, Phase diagram and dielectric properties of MA$_{1-x}$FA$_{x}$PbI$_{3}$, \textcolor{blue}{\href{https://pubs.acs.org/doi/full/10.1021/acsenergylett.9b01291}{ACS Energy Lett. {\bf 4 }, 2045 (2019)}}.

\bibitem{Fabini-emphanisis}D. H. Fabini, G. Laurita, J. S. Bechtel, C. C. Stoumpos, H. A. Evans, A. G. Kontos, Y. S. Raptis, P. Falaras, A. Van der Ven, M. G. Kanatzidis, and R. Seshadri, Dynamic stereochemical activity of the Sn$^{2+}$ lone pair in perovskite CsSnBr$_3$,  \textcolor{blue}{\href{https://pubs.acs.org/doi/full/10.1021/jacs.6b06287} {J. Am. Chem. Soc. {\bf 138}, 11820 (2016)}}.

\bibitem{Debasmita_JACS2023}D. Pariari, S. Mehta, S. Mandal, A. Mahata, T. Pramanik, S. Kamilya, A. Vidhan, T. N. Guru Row, P. K. Santra, S. K. Sarkar, F. De Angelis, A. Mondal, and D. D. Sarma, Realizing the lowest bandgap and exciton binding energy in a two-dimensional lead halide system, \textcolor{blue}{\href{https://pubs.acs.org/doi/full/10.1021/jacs.3c03300}{J. Am. Chem. Soc., {\bf 145}, 15896 (2023)}}.

\bibitem{Debasmita_ACSEL2024}D. Pariari, P. Acharyya, A. Sinha, A. Mohanty, S. Sett, N. K. Gill, A. Ghosh, U. V. Waghmare, K. Biswas, and D. D. Sarma,  Non-monotonic thermal conductivity of FA$_x$MA$_{1-x}$PbI$_3$ achieving ultralow values: The role of anharmonic low energy rotation of organic moieties, \textcolor{blue}{\href{https://pubs.acs.org/doi/10.1021/acsenergylett.4c00047}{ACS Energy Lett. {\bf 9}, 2128 (2024)}}.

\bibitem{Kang_JPCL_2021}K. Kang, W. Hu, and X. Tang, Halide perovskites for resistive switching memory, \textcolor{blue}{ \href{https://pubs.acs.org/doi/full/10.1021/acs.jpclett.1c03408}{J. Phys. Chem. Lett  {\bf 12}, 11673 (2021)}}.

\bibitem{Lin-Lai_Dou}J. Lin, M. Lai, L. Dou, C. S. Kley, H. Chen, F. Peng, J. Sun, D. Lu, S. A. Hawks, C. Xie, F. Cui, A. P. Alivisatos, D. T. Limmer, and P. Yang, Thermochromic halide perovskite solar cells, \textcolor{blue}{\href{https://www.nature.com/articles/s41563-017-0006-0}{Nat. Mat. {\bf 17}, 261 (2018)}}. 

\bibitem{Lin_VO2_PRL2022}L. Jin, Y. Shi, F. I. Allen, L.-Q. Chen, and J. Wu, Probing the critical nucleus size in the metal-insulator phase transition of  VO$_2$, \textcolor{blue}{\href{http://link.aps.org/doi/10.1103/PhysRevLett.129.245701}{Phys. Rev. Lett. {\bf 129}, 245701 (2022).}}

\bibitem{Ortin_Planes_Delaey}J. Ortin, A. Planes, and L. Delaey, \textcolor{blue}{\href{https://doi.org/10.1016/B978-012480874-4/50023-3}{Hysteresis in shape-memory materials}}, The Science of Hysteresis, edited by G. Bertotti and I. Mayergoyz (Elsevier, Oxford, U. K., 2005), Vol. III, p. 467.

\bibitem{Bhadeshia}H. K. D. H. Bhadeshia and C.M. Wayman, \textcolor{blue}{\href{http://dx.doi.org/10.1016/B978-0-444-53770-6.00009-5}{Phase transformations: Nondiffusive}} in \textit{Physical Metallurgy}, 5th ed., edited by D. E. Laughlin and K. Hono (Elsevier, Amsterdam, 2014), pp 1021-1072.

\bibitem{Nishiyama}Z. Nishiyama, \textcolor{blue}{\href{https://www.elsevier.com/books/martensitic-transformation/nishiyama/978-0-12-519850-9}{{\it Martensitic Transformation}}} (Academic Press, New York 1978).

\bibitem{Nandi}S. K. Nandi, G. Biroli, and G. Tarjus, Spinodals with disorder: From avalanches in random magnets to glassy dynamics, \textcolor{blue}{\href{https://journals.aps.org/prl/abstract/10.1103/PhysRevLett.116.145701}{Phys. Rev. Lett. {\bf 116}, 145701 (2016).}}

\bibitem{Chandni}U. Chandni, A. Ghosh, H.S. Vijaya, and S. Mohan, Criticality of tuning in athermal phase transitions, \textcolor{blue}{\href{http://link.aps.org/doi/10.1103/PhysRevLett.102.025701}{Phys. Rev. Lett. {\bf 102}, 025701 (2009).}}

\bibitem{Perez-Reche_PRL2001}F. J. P\'erez-Reche, E. Vives, L. Ma\~nosa, and A. Planes, Athermal Character of Structural Phase Transitions,  \textcolor{blue}{\href{https://journals.aps.org/prl/abstract/10.1103/PhysRevLett.87.195701}{Phys. Rev. Lett. {\bf 87}, 195701 (2001)}}.

\bibitem{Suppl}See Supplemental Material, which includes Refs.  \onlinecite{Mayergoyz_PRL1986, Perkovic_Sethna, Muller_Viscosity, Yeshurun_RMP, Balankin_PRE2011, Eisenbach_ElectronGlass, Rankine_1907}, for the details of the growth, the structural characterization of MAPbI$_3$ samples and more data including the return-point memory behavior seen from the x-ray measurements and the Preisach analysis of the hysterons maps.   

\bibitem{Mayergoyz_PRL1986}I. D. Mayergoyz, Mathematical models of hysteresis, \textcolor{blue}{\href{https://doi.org/10.1103/PhysRevLett.56.1518}{Phys. Rev. Lett. {\bf 56}, 1518 (1986)}}.

\bibitem{Perkovic_Sethna}O. Perkovi\'c and J. P. Sethna, Improved magnetic information storage using return-point memory, \textcolor{blue}{\href{https://doi.org/10.1063/1.364088}{J. Appl. Phys. {\bf 81}, 1590 (1997)}}.

\bibitem{Muller_Viscosity}K. -H. M\"uller, Magnetic viscosity, \textcolor{blue}{ \href{https://www.sciencedirect.com/science/article/abs/pii/B008043152600869X}{The Encyclopedia of Materials: Science and Technology}}, edited by K. H. J. Buschow, R. W. Cahn,  M. C. Flemings, B. Ilschner, E. Kramer, and S. Mahajan (Elsevier Science Ltd., New York, 2001), pp. 4997--5004.

\bibitem{Yeshurun_RMP}Y. Yeshurun, A. P. Malozemoff, and A. Shaulov, Magnetic relaxation in high-temperature superconductors, \textcolor{blue}{ \href{https://doi.org/10.1103/RevModPhys.68.911}{Rev. Mod. Phys. {\bf 68}, 911 (1996)}}.

\bibitem{Balankin_PRE2011} A. S. Balankin, O. S. Huerta, F. Hern\'andez M\'endez, and  J. Patino Ortiz, Slow dynamics of stress and strain relaxation in randomly crumpled elasto-plastic sheets, \textcolor{blue}{\href{https://link.aps.org/doi/10.1103/PhysRevE.84.021118}{Phys. Rev. E {\bf 84}, 021118 (2011)}}.

\bibitem{Eisenbach_ElectronGlass}A. Eisenbach, T. Havdala, J. Delahaye, T. Grenet, A. Amir, and A. Frydman, Glassy dynamics in disordered electronic systems reveal striking thermal memory effects, \textcolor{blue}{\href{http://dx.doi.org/10.1103/PhysRevLett.117.116601}{Phys. Rev. Lett. {\bf 117}, 116601 (2016)}}; \textcolor{blue}{\href{https://doi.org/10.1103/PhysRevLett.117.139901}{Phys. Rev. Lett. {\bf 117}, 139901(E) (2016)}}.

\bibitem{Rankine_1907}A. O. Rankine, The behaviour of over-strained materials, \textcolor{blue}{\href{https://www.jstor.org/stable/43769106}{Sci. Prog. Twentieth Century (1906-1916)}}, {\bf 1}, 465 (1907).

\bibitem{fn_defects}The PL intensity shows a small logarithmic increase with time following due to the mechanical accommodation associated with the structural transition and the healing of defects. This is discussed in the Supplementary Material. The spectral shape, which is a more direct measure of the phase fraction, is arrested to a much higher degree.

\bibitem{Sethna_RPP}J. P. Sethna, K. Dahmen, S. Kartha, J. A. Krumhansl, B. W. Roberts, and J. D. Shore, Hysteresis and hierarchies: Dynamics of disorder-driven first-order phase transformations, \textcolor{blue}{\href{https://journals.aps.org/prl/abstract/10.1103/PhysRevLett.70.3347}{Phys. Rev. Lett. {\bf 70}, 3347 (1993)}}.

\bibitem{Debenedetti}P. G. Debenedetti, \textcolor{blue}{\href{https://press.princeton.edu/books/hardcover/9780691085951/metastable-liquids}{{\it Metastable Liquids: Concepts and Principles}}} (Princeton University Press, Princeton, NJ 1997).

\bibitem{Wang_Ong} Z. Z. Wang and N. P. Ong, Disequilibration of the pinned charge-density-wave state by slight changes in temperature, \textcolor{blue}{\href{https://journals.aps.org/prb/abstract/10.1103/PhysRevB.34.5967}{Phys. Rev. B {\bf 34, 5967} (1986)}}; N.P. Ong and Z.Z. Wang, \textcolor{blue}{\href{https://link.springer.com/chapter/10.1007/978-3-642-83033-4_40}{Non-equilibrium behavior, hysteresis and condensate quakes in the pinned charge density wave}} in Nonlinearity in Condensed Matter--Proceedings of the Sixth Annual Conference, Center for Nonlinear Studies Los Alamos, New Mexico, 1986,  edited by A. R. Bishop, D. K. Campbell, P. Kumar, and S. E. Trullinger (Springer, Berlin, 1987), p. 350.

\bibitem{Gottschall}T. Gottschall, A. Gr\`acia-Condal, M. Fries, A. Taubel, L. Pfeuffer, L. Ma\~nosa, A. Planes, K. P. Skokov, and O. Gutfleisch, A multicaloric cooling cycle that exploits thermal hysteresis,  \textcolor{blue}{\href{https://www.nature.com/articles/s41563-018-0166-6}{Nat. Mater. {\bf 17}, 929 (2018)}}.

\bibitem{Olemskoi-Katsnelson}O. I. Olemskoi and A. A. Katsnelson, Structural transformations far off equilibrium, \textcolor{blue}{\href{https://doi.org/10.15407/ufm.03.01.001}{Usp. Fiz. Met. \bf {3},  1 (2002)}} (in Russian).

\bibitem{Satyaki_PRL}S. Kundu, T. Bar, R. K. Nayak, and B. Bansal, Critical slowing down at the abrupt Mott transition: When the first-order phase transition becomes zeroth order and looks like second order, 	 \textcolor{blue}{\href{https://journals.aps.org/prl/abstract/10.1103/PhysRevLett.124.095703}{Phy. Rev. Lett. {\bf 124}, 095703 (2020)}}.

\bibitem{Bhattacharya}K. Bhattacharya, \textcolor{blue}{\href{https://academic.oup.com/book/54729}{Microstructure of Martensite: Why It Forms and How It Gives Rise to the Shape Memory Effect}} (Oxford University Press, New York, 2003). 

\bibitem{Zhang_James_Muller}Z. Zhang, R. D. James, and S. M\"uller, Energy barriers and hysteresis in martensitic phase transformations, \textcolor{blue}{\href{https://doi.org/10.1016/j.actamat.2009.05.034}{Acta Mater. {\bf 57}, 4332 (2009)}}.

\bibitem{Latella_Madrid_Ruffo}I. Latella, A. P\'erez-Madrid, A. Campa, L. Casetti, and S. Ruffo, Thermodynamics of nonadditive systems, \textcolor{blue}{\href{https://doi.org/10.1103/PhysRevLett.114.230601}{Phys. Rev. Lett. {\bf 114}, 230601 (2015)}}.

\bibitem{Gagne_Gould_Klein}C. J. Gagne, H. Gould, W. Klein, T. Lookman, and A. Saxena, Simulations of spinodal nucleation in systems with elastic interactions, \textcolor{blue}{\href{https://doi.org/10.1103/PhysRevLett.95.095701}{Phys. Rev. Lett. {\bf 95}, 095701 (2005)}}.

\bibitem{Pazmandi_PRL1999}F. P\'azm\'andi, G. Zar\'and, and G. T. Zim\'anyi, Self-organized criticality in the hysteresis of the Sherrington-Kirkpatrick model, \textcolor{blue}{\href{https://link.aps.org/doi/10.1103/PhysRevLett.83.1034}{Phys. Rev. Lett. {\bf 83}, 1034 (1999)}}.

\bibitem{ladyman}J. Ladyman and K. Wiesner, {\it What Is a Complex System?} (Yale University Press, New Haven 2020).

\end{thebibliography}
\end{document}